\documentclass[sigplan,screen]{acmart}
\AtBeginDocument{%
  }

\copyrightyear{2025}
\acmYear{2025}
\setcopyright{cc}
\setcctype{by-nc}
\acmConference[ASPLOS '25] {Proceedings of the 30th ACM International Conference on Architectural Support for Programming Languages and Operating Systems, Volume 2}{March 30--April 3, 2025}{Rotterdam, Netherlands.}
\acmBooktitle{Proceedings of the 30th ACM International Conference on Architectural Support for Programming Languages and Operating Systems, Volume 2 (ASPLOS '25), March 30--April 3, 2025, Rotterdam, Netherlands}
\acmDOI{10.1145/3676641.3716256}
\acmISBN{979-8-4007-1079-7/2025/03}

\usepackage[]{hyperref}
\usepackage{enumitem}
\usepackage{algorithmicx}
\usepackage{algpseudocode}

\usepackage{braket}
\usepackage{makecell}
\usepackage{algorithm}

\begin{document}

%%
%% The "title" command has an optional parameter,
%% allowing the author to define a "short title" to be used in page headers.
%\title{Fat-Tree QRAM: A High-Bandwidth Shared Quantum Random Access Memory for Parallel Queries}
\title[Fat-Tree QRAM: A High-Bandwidth Shared Quantum Random Access Memory \\ for Parallel Queries]{Fat-Tree QRAM: A High-Bandwidth Shared Quantum Random Access Memory for Parallel Queries}

%%
%% The "author" command and its associated commands are used to define
%% the authors and their affiliations.
%% Of note is the shared affiliation of the first two authors, and the
%% "authornote" and "authornotemark" commands
%% used to denote shared contribution to the research.
\author{Shifan Xu}
% \affiliation{%
%   \institution{Yale University}
%   %\department{Departments of Physics and Applied Physics}
%   \city{New Haven}
%   \state{CT}
%   \postcode{06511}
%   \country{USA}}
\orcid{0009-0005-9103-228X}
\affiliation{%
  \department{Yale Quantum Institute}
  \institution{Yale University}  
  \city{New Haven}
  \state{CT}
  \postcode{06520-8263}
  \country{USA}}
\email{shifan.xu@yale.edu}

\author{Alvin Lu}
\orcid{0009-0004-5010-6871}
\affiliation{%
  \department{Yale Quantum Institute}
  \institution{Yale University}
  \city{New Haven}
  \state{CT}
  \postcode{06520-8263}
  \country{USA}}
\email{alvin.lu@yale.edu}

\author{Yongshan Ding}
\orcid{0000-0002-2338-1315}
\affiliation{%
  \department{Yale Quantum Institute}
  \institution{Yale University}
  \city{New Haven}
  \state{CT}
  \postcode{06520-8263}
  \country{USA}}
\email{yongshan.ding@yale.edu}

%%
%% By default, the full list of authors will be used in the page
%% headers. Often, this list is too long, and will overlap
%% other information printed in the page headers. This command allows
%% the author to define a more concise list
%% of authors' names for this purpose.

\renewcommand{\shortauthors}{Shifan Xu, Alvin Lu, \& Yongshan Ding}

%%
%% The abstract is a short summary of the work to be presented in the
%% article.
\begin{abstract}
Quantum Random Access Memory (QRAM) is a crucial architectural component for querying classical or quantum data in superposition, enabling algorithms with wide-ranging applications in quantum arithmetic, quantum chemistry, machine learning, and quantum cryptography. In this work, we introduce Fat-Tree QRAM, a novel query architecture capable of pipelining multiple quantum queries simultaneously while maintaining desirable scalings in query speed and fidelity. Specifically, Fat-Tree QRAM performs 
$O(\log (N))$ independent queries in $O(\log (N))$ time using $O(N)$ qubits, offering immense parallelism benefits over traditional QRAM architectures. To demonstrate its experimental feasibility, we propose modular and on-chip implementations of Fat-Tree QRAM based on superconducting circuits and analyze their performance and fidelity under realistic parameters. Furthermore, a query scheduling protocol is presented to maximize hardware utilization and access the underlying data at an optimal rate. These results suggest that Fat-Tree QRAM is an attractive architecture in a shared memory system for practical quantum computing.
\end{abstract}

%%
%% The code below is generated by the tool at http://dl.acm.org/ccs.cfm.
%% Please copy and paste the code instead of the example below.
%%


%%
%% Keywords. The author(s) should pick words that accurately describe
%% the work being presented. Separate the keywords with commas.
%% A "teaser" image appears between the author and affiliation
%% information and the body of the document, and typically spans the
%% page.

% \received{20 February 2007}
% \received[revised]{12 March 2009}
% \received[accepted]{5 June 2009}

%%
%% This command processes the author and affiliation and title
%% information and builds the first part of the formatted document.
\maketitle

\section{Introduction}

Many quantum algorithms for solving classically intractable problems assume that a large classical or quantum memory can be queried in superposition. Bucket-Brigade Quantum Random Access Memory (BB QRAM)~\cite{giovannetti2008quantum} is a promising candidate for realizing such queries efficiently, achieving desirable (poly-)logarithmic scalings in query latency and infidelity relative to the memory size~\cite{hann2021practicality}.  Recent resource estimates have revealed that QRAM's utility in quantum algorithms varies depending on the input data size and algorithmic speedup. For instance, a quadratic speedup in Grover's algorithm~\cite{grover1996fast} for database search is insufficient to realize a practical quantum advantage~\cite{jaques2023qram, hoefler2023disentangling}. However, QRAM remains central to enabling quantum advantages in many algorithms like the qubitization algorithm for chemistry simulation~\cite{lee2021even, berry2019qubitization, van2020convex}, Harrow-Hassidim-Lloyd algorithm for solving systems of equations and machine learning~\cite{biamonte2017quantum, harrow2009quantum}, and variants of Shor's algorithm for prime factorization~\cite{gidney2021factor, shor1994algorithms}. 

Running these quantum algorithms is challenging due to their demanding resource requirements, including large numbers of qubits with long coherence times. Specifically, these algorithms are inherently sequential---they make serial QRAM queries and consequently require deep circuits. These challenges can be alleviated through a parallel processing approach. Motivated by the ubiquitous use of parallelism in classical computation, numerous parallel quantum algorithms have recently emerged. Examples include distributed variational quantum eigensolver (VQE) \cite{niu2023parameter}, distributed Shor's algorithm \cite{meter2006architecture}, distributed quantum phase estimation (QPE) \cite{liu2021distributed, ang2022architectures}, parallel quantum walk \cite{zhang2024parallel}, and parallel quantum signal processing (QSP) \cite{martyn2024parallel}. The success of many parallel algorithms critically depends on high-bandwidth QRAM capable of supporting simultaneous queries.

\begin{figure}
    \centering
    \includegraphics[width=\linewidth]{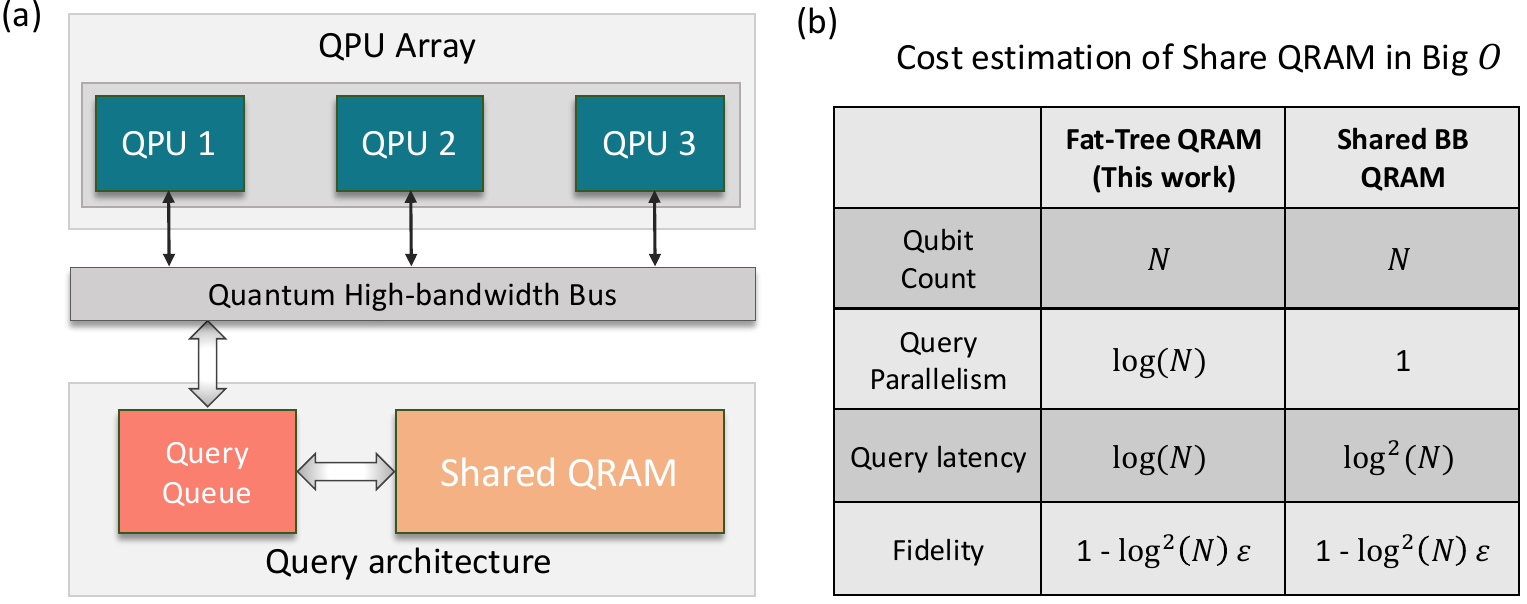} % Placeholder for now
    \caption{(a) Architectural schematics of a shared QRAM that is concurrently accessed by multiple QPUs. (b) Cost comparison between Fat-Tree and Bucket-Brigade (BB) QRAMs for executing $O(\log(N))$ independent queries. The proposed Fat-Tree QRAM allows $O(\log(N))$ queries to be executed in parallel while maintaining desirable asymptotic scalings, including $O(N)$ qubit count, $O(\log(N))$ total latency (i.e., circuit depth), and $O(\log^2(N) \varepsilon)$ infidelity.}
    \label{fig:motivation}
\end{figure}

In tandem with algorithmic advances, tremendous hardware progress has been made towards realizing QRAM. Multiple platforms have successfully demonstrated fast and high-fidelity controlled-SWAP (\texttt{CSWAP}) gates, a critical native operation in QRAM~\cite{xue2023hybrid,leger2024implementation,gao2019entanglement}. Experimental QRAM prototypes have been proposed based on quantum optics~\cite{jiang2019experimental}, Rydberg atoms \cite{patton2013ultrafast}, photonics \cite{chen2021teleqram}, and circuit quantum acoustodynamics \cite{hann2019hardware}, and superconducting cavities \cite{weiss2024quantum}. Yet, one of the most substantial limitations of QRAM is the large number of qubits required for practically relevant problems, typically $O(N)$ qubits for a size-$N$ memory. This issue can be mitigated by a \emph{shared memory} approach, where multiple quantum processing units (QPUs) share QRAM resources to improve utilization, as illustrated schematically in Fig.~\ref{fig:motivation}(a). For example, recent proposals for quantum data centers (QDC) \cite{liu2022quantum,liu2023data,liu2024quantum} have highlighted the utility of a shared QRAM system for quantum applications including multi-party private communication and quantum sensing \cite{liu2022quantum}. This shared QRAM model~\cite{alexeev2021quantum} also aligns well with the technology trends towards distributed or multi-core quantum computing, where multiple users can access shared quantum systems via cloud. Meanwhile, the emerging modular approach of building complex quantum systems from smaller modules also provides hardware support for such large-scale quantum computing architectures \cite{monroe2014large,bombin2021interleaving,krutyanskiy2023entanglement}. However, existing QRAM architectures, such as the BB QRAM, have extremely poor performances under contention. That is, a single query occupies all $O(N)$ quantum routers for the entire duration of the query. Consequently, queries must be queued and executed sequentially.

In this work, we introduce a novel shared QRAM architecture that pipelines multiple independent queries simultaneously while preserving the qubit number and query fidelity scalings of a BB QRAM. We term this design ``Fat-Tree QRAM,'' as the organization of the quantum routers resembles a Fat-Tree \cite{leiserson1985fat} that is commonly seen in classical computing and networking systems. 

\begin{itemize}
    \item Fat-Tree QRAM architecture pipelines $O(\log(N))$ independent queries to a size-$N$ memory in $O(\log(N))$ time using $O(N)$ qubits (Fig. ~\ref{fig:motivation}(b)). This approach provides a scalable path towards building a hardware-efficient, high-bandwidth quantum shared memory system. 
    \item We consider both modular and on-chip implementations of the Fat-Tree QRAM architecture using superconducting cavities. While our QRAM design can be generalized to any technology platform that supports native \texttt{CSWAP} operations, we demonstrate that Fat-Tree QRAM can be efficiently implemented despite restrictive connectivity constraints in superconducting platforms.
    \item We analyze the optimal query scheduling/pipelining protocol that resolves resource contention and maximizes utilization and throughput for parallel queries. We discuss the benefits of such parallelism in the context of parallel quantum algorithms and parallel execution of multiple quantum algorithms. 
    \item Of great interests from an experimental standpoint are the new metrics we introduced to benchmark shared QRAM architectures, including QRAM bandwidth, space-time volume per query, hardware utilization, and memory access rate. 
\end{itemize}

Our paper is organized as follows. Sec.~\ref{sec:background} reviews current noisy intermediate-scale quantum (NISQ) machines and state-of-the-art quantum random access memory architectures. In Sec.~\ref{sec:app} and Sec.~\ref{sec:arch}, we explore quantum shared memory systems by introducing the hardware architecture of Fat-Tree QRAM with detailed implementations based on superconducting circuits. In Sec.~\ref{sec:schedule}, we provide a scheduling protocol to maximize the utilization of Fat-Tree QRAM. In Sec.~\ref{sec:eval} and Sec.~\ref{sec:results}, we evaluate the performance of the Fat-Tree QRAM for both real-world parallel quantum algorithms and synthetic algorithms. We conclude with a brief discussion on the implication of these results for large-scale quantum computing.

\section{Background}\label{sec:background}
\begin{figure*}[t]
    \centering
    \includegraphics[width=1.0\textwidth]{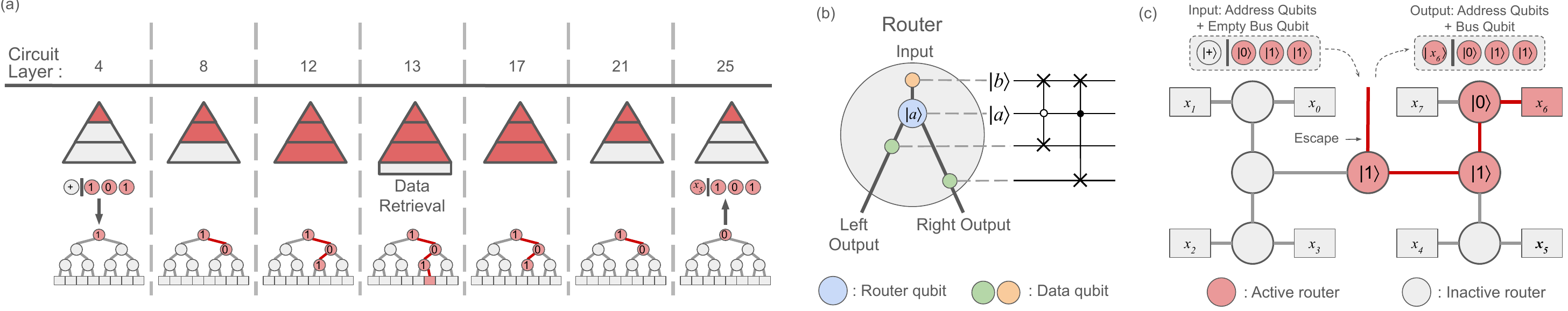}
    \caption{(a) Querying a Bucket-Brigade (BB) QRAM with capacity $N=8$ takes 25 circuit layers. A detailed step-by-step procedure can be found in Appendix~\ref{subsec:BBdetail}. The circuit layer number indicates the \emph{finishing} time of each stage. (b) Each quantum router in the BB QRAM involves \texttt{CSWAP} operations between the router qubit and the data qubits. (c) H-tree layout of a BB QRAM. Quantum routers are organized in a binary tree structure, where classical data are located at the leaves. Dashed lines indicate external address and bus qubits that are used to query the QRAM. Figure (c) represents the QRAM state after all address qubits have been loaded, corresponding to circuit layer 15 in (a).} 
    \label{fig:bg}
\end{figure*}

\subsection{Emerging Quantum Hardware and Software} 

Quantum algorithms have been shown to provide polynomial or super-polynomial speedups against their best-known classical counterparts on special computational tasks, ranging from quantum chemistry simulations~\cite{lee2021even, berry2019qubitization, van2020convex} to quantum cryptography~\cite{gidney2021factor, shor1994algorithms}. Many of these algorithms, however, rely on the existence of a quantum random access memory (QRAM) device to efficiently query classical or quantum data in superposition, coupled with a quantum processing unit (QPU) with sufficient system size to process the queried data efficiently and fault tolerantly. For example, a classically intractable problem of scientific or industry interests is expected to require hundreds of thousands of high-fidelity qubits \cite{gidney2021factor, beverland2022assessing}.

In recent years, many physical architecture platforms have demonstrated high-quality control over tens or hundreds of qubits. Due to various constraints including connectivity, power, and wiring, it is challenging to scale up these systems as a single monolithic quantum processor. These constraints can be mitigated by a \emph{modular} approach, where small units of special-purpose devices are linked together to form a single quantum processor~\cite{meter2008arithmetic, monroe2014large, nickerson2014freely, alexeev2021quantum}. This modular approach (both in hardware and software) can also simplify manufacturing, control, and maintenance. For example, in superconducting quantum computers, tunable couplers can be connected to bendable cryogenic microwave cables to mediate cross-chip interactions between remote qubits from separate modules. The flexibility of the microwave links allows the system's connectivity to extend beyond planar layouts. More compact on-chip designs have also been demonstrated, but typically have stricter topology constraints to avoid crossing inter-connecting wires. Multi-layer die stacks can alleviate this challenge but require vertical connection using Through-Silicon-Vias (TSVs) \cite{yost2020solid,gambino2015overview,rosenberg20173d,brecht2016multilayer}.

The rapid increment of available qubits and advances in scalable architectures also facilitate the development of quantum algorithms. For example, running multiple algorithms in shared hardware makes it possible for IBM's 1000+ qubit machine to increase hardware utilization~\cite{gambetta2020ibm}. Meanwhile, many quantum algorithms benefit from the novel distributed or multi-core quantum computing architectures, dramatically improving resource efficiency and overall performance~\cite{wu2023qucomm,caleffi2022distributed}.

\subsection{Quantum Queries}
A quantum random access memory implements a quantum query by accessing (classical) memory at multiple addresses in superposition. It realizes the following unitary operation: 
\begin{align}
\sum_{i=0}^{N-1}\alpha_i\ket{i}_{\text{A}}\ket{0}_{\text{B}} &\xrightarrow[]{\text{Query}} \sum_{i=0}^{N-1}\alpha_i\ket{i}_{\text{A}}\ket{x_i}_{\text{B}}\label{eq:qram}
\end{align}
where $\ket{\cdot}_{\text{A}}$ ($\ket{\cdot}_{\text{B}}$) is the address (bus) qubit register storing the input (output), $x_i$ is the data value stored at address $i$, and $\alpha_i$ is the superposition amplitude of address $i$. $N$ is the size of the memory (or QRAM capacity). The number of address and bus qubits, $|A|$ and $|B|$ respectively, are termed the \emph{address width} and \emph{bus width}. For the remainder of the paper, we will assume $|A|=\log(N)$ and $|B|=1$. 

\subsubsection{Overview of Bucket-Brigade QRAM}

We consider BB QRAM, one of the leading quantum query architectures, proposed by Giovannetti et al. in 2008 \cite{giovannetti2008quantum, giovannetti2008architectures}. BB QRAM implements a quantum query to a memory of size $N$ in $O(\log(N))$ time (i.e., circuit layer \cite{amico2023defining}, which is defined as one logical circuit step where all quantum gates inside the same
layer are executed in parallel). BB QRAM is also proven to exhibit superior noise resilience than other architectures, including Fanout QRAM \cite{nielsen2010quantum} and Select-Swap QRAM \cite{low2024trading}.  

The basic building block of a BB QRAM is a \emph{quantum router}. Shown in Fig.~\ref{fig:bg}(b), a quantum router consists of two \texttt{CSWAP} gates acting on four qubits. The two \texttt{CSWAP} gates route an input qubit to either the left or right output qubits in a superposition based on the quantum state of the router qubit which takes one of three states: $\ket{W}$ inactive ``wait'' state routes trivially, $\ket{0}$ routes left, and $\ket{1}$ routes right. BB QRAM recursively concatenates quantum routers initialized to $\ket{W}$ in a binary tree structure. Fig.~\ref{fig:bg}(c) shows BB QRAM in a 2D H-Tree layout \cite{giovannetti2008architectures, xu2023systems}.

\subsubsection{Query Procedure in BB QRAM}

We define four main operations for BB QRAM routers: \texttt{LOAD} (L) qubit through escape, \texttt{TRANSPORT} (T) to next router, \texttt{ROUTE} (R) in current router, and \texttt{STORE} (S) into router qubit. Detailed definitions for each can be found in Appendix~\ref{subsec:BBdetail}.

Using the four operations, BB QRAM realizes quantum queries in three stages: \emph{address loading}, \emph{data retrieval}, and \emph{address unloading}. In address loading, each $i^{\text{th}}$ address qubit is \texttt{LOAD}ed through the escape and then routed to the $i^{\text{th}}$ level of the tree by a series of alternating \texttt{ROUTE} and \texttt{TRANSPORT} operations. The specific path is controlled by the previously routed address qubits stored in higher levels of the tree. Once an address qubit is at the right level, it is \texttt{STORE}d into the routers. Note that each address qubit can be loaded and begin routing before the previous address qubit has been stored since the beginning of the path is independent of the last address qubit; this ``bit-level pipelining'' (to distinguish from ``query-level pipelining'' introduced by this paper in later sections) reduces the total latency through quantum parallelism. After address loading, the fully loaded QRAM stores a superposition of different addresses, where each address activates a distinct root-to-leaf path that is unentangled with the other routers in the $\ket{W}$ state (important for maintaining fidelity).

In the data retrieval stage, the bus qubit is routed to the leaves of the QRAM tree in a procedure similar to address loading (also as part of the ``bit-level pipeline'', i.e., loaded before the last address qubit is stored). All classical memory are copied in parallel to modify the ``delocalized bus qubit'' at the leaves of the QRAM tree. Finally, the bus qubit is routed out of the tree, and the routers are reverted to an all-$\ket{W}$ state through uncomputation, which follows the same steps as address loading but in reverse. A step-by-step description and instruction set can be found in Appendix~\ref{subsec:BBdetail}.

The inherent parallelism in executing both quantum gates and classical queries ensures the BB QRAM has an $O(\log(N))$ latency for address loading (4 circuit layers for storing each address qubit and routing the bus), an $O(1)$ latency for data retrieval (though also a single circuit layer, it is much faster than other gates in practice), and an overall $O(\log(N))$ query latency. We provide a visual description of the query procedure of BB QRAM in Fig.~\ref{fig:bg}(a) and a more detailed version in Appendix~\ref{subsec:BBdetail}.

It has been shown that BB QRAM has intrinsic noise resilience, due to limited entanglement among different paths and restricted propagation of errors. The infidelity of a query is proven to be upper bounded by $O(\epsilon\log^2(N))$, where $\epsilon$ is the error rate of each operation and $N$ is the size of the memory \cite{hann}. Such superior infidelity scaling makes BB QRAM a particularly attractive candidate for implementation before the era of fault tolerance.

\section{Challenges and Motivation}
\label{sec:app}

Despite its speed and fidelity advantages for completing a single query, BB QRAM is not capable of processing multiple queries in parallel. This limitation is intrinsic to the binary tree structure of the QRAM architecture. For example, the $0^{\text{th}}$ address qubit is routed into the tree and occupies the root node for the entire duration of the query. In a binary tree structure, the root node serves as the sole escape route (i.e., external interface) through which every address qubit must pass. Consequently, all queries must be queued and executed sequentially.

In a shared memory system, as illustrated in Fig.~\ref{fig:motivation}, a BB QRAM inevitably leads to \emph{resource contention}. When $p$ parallel processes attempt to query the shared memory, BB QRAM must to execute them sequentially. This lack of query parallelism leads to a total query latency of $O(p \log(N))$, potentially causing a slowdown in quantum algorithms. Motivated by advancements in parallel computing and networking in classical literature, we propose an alternative router-based QRAM architecture based on a Fat-Tree structure, similar to the Fat-Tree network initially proposed by Charles Leiserson in 1985 \cite{leiserson1985fat}. 
Indeed, Fat-Tree QRAM routes differently than a classical Fat-Tree network~\cite{leiserson1985fat}, despite their similarity in geometry. A useful conceptual picture is that qubits are routed from root to leaf in QRAM, as opposed to communicating among leaf memory cells. With only a moderate (i.e., small constant factor) increase in the number of qubits in quantum routers at the higher levels of the tree, we can pipeline multiple queries simultaneously, offering immense parallelism benefits to a shared memory system.

\section{Shared QRAM Architecture}
\label{sec:arch}

\begin{figure}[t]
         \centering
         \includegraphics[width=0.48\textwidth]{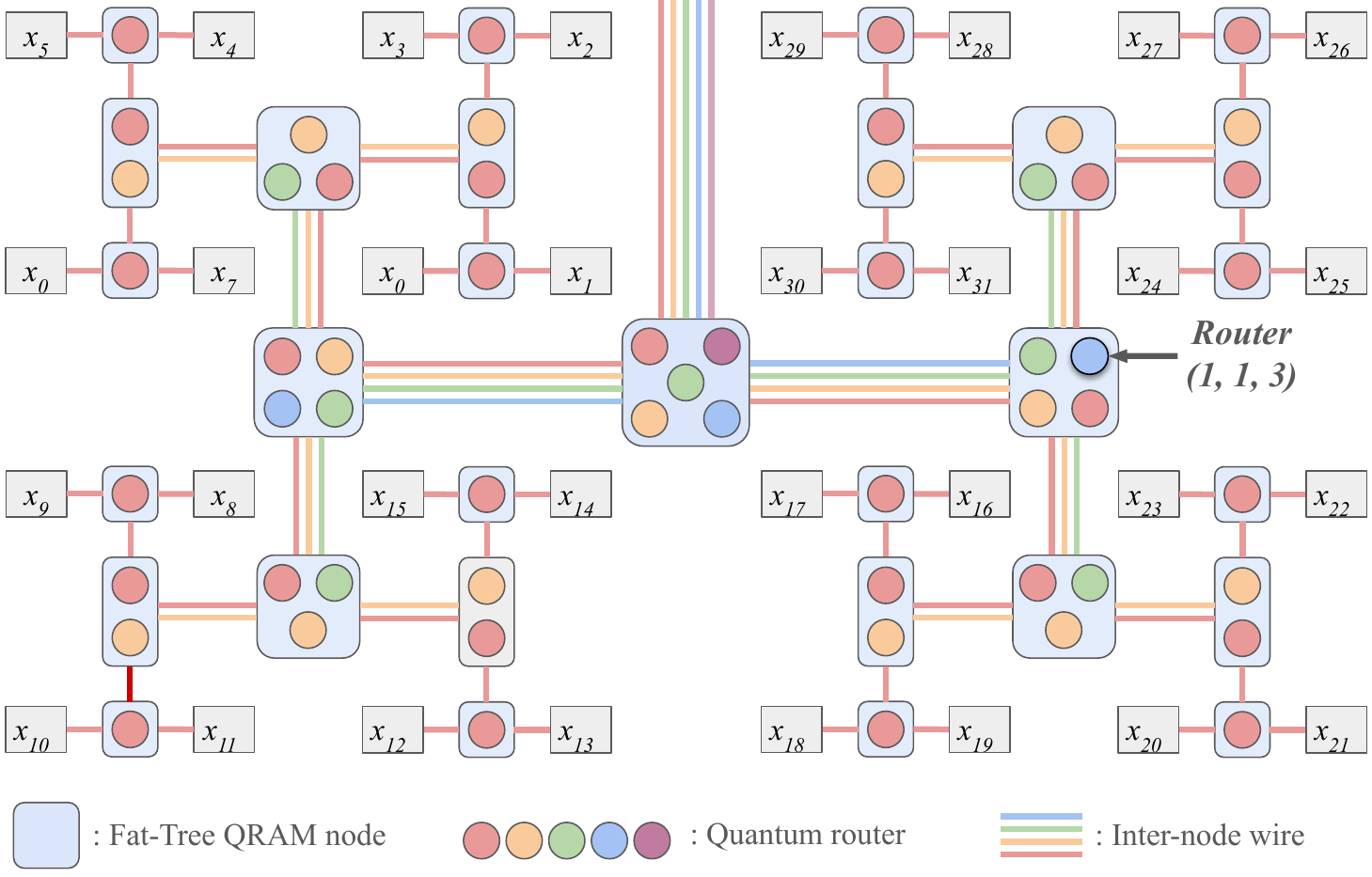}
         \caption{Layout of a Fat-Tree QRAM with capacity $N=32$ (Similar H-tree layout for BB QRAM appeared in \cite{xu2023systems}). Classical data are located at the leaves and the internal nodes contain multiplexed quantum routers. Colors of the routers and wires are used to indicate connection. The size of an internal node (i.e., number of qubits) increases \emph{linearly} as we go up the tree. 
         }
         \label{fig:fat}
\end{figure}

We introduce the new Fat-Tree QRAM architecture in the following three subsections: Sec.~\ref{subsec:fattree} describes a Fat-Tree architecture to increase the query parallelism. Sec.~\ref{subsec:nodes} provides both modular and on-chip hardware demonstrations of Fat-Tree QRAM using well-established techniques in NISQ systems. Finally, operations within the architecture and the query pipeline diagram are presented in Sec.~\ref{subsec:operation}.

\subsection{Fat-Tree Architecture}
\label{subsec:fattree}

Fat-Tree QRAM is built on a complete binary tree, where $N$ leaf nodes connect to size-$N$ classical memory. Each tree level consists of quantum routers, with the $i^{\text{th}}$ level (indexed from 0) containing $2^i$ routers, as in BB QRAM. To pipeline $n=\log(N)$ queries with address width $n$, routers at level $i$ are duplicated $n-i-1$ times. Thus, Fat-Tree QRAM adopts a 2D H-tree layout similar to BB QRAM, replacing each router at level $i$ with a Fat-Tree node containing $n-i$ routers. Fig.~\ref{fig:fat} illustrates this structure, where circles inside square nodes denote quantum routers. 

We use a 3-tuple $(i,j,k)$ to index routers and qubits: $i\in [0, n-1]$ represents the level, $j\in [0, 2^i-1]$ denotes the node index, and $k\in [0, n-i-1]$ identifies the router copy in node $(i,j)$. In BB QRAM, routers correspond to $(i,j,n-1)$. The parameter $k$ determines multiplexing, extending BB QRAM into the Fat-Tree model. Individual qubits within each router are categorized as input, router, and L/R output qubits (e.g., the input qubit of router $(1,1,3)$ in Fig.~\ref{fig:fat}).

Increased router duplication raises inter-node connectivity. BB QRAM links parent-child nodes with a single wire (Fig.~\ref{fig:bg}), whereas Fat-Tree QRAM connects nodes with $k$ wires per node. Starting with $n$ wires at the root, the count decreases by 1 per level until reaching a single wire at the leaves, matching BB QRAM. This enhanced connectivity enables higher-bandwidth inter-node communication and multiple parallel gates between nodes.

As discussed in Sec.~\ref{sec:app}, the resource overhead from duplicating higher BB QRAM levels remains moderate. The qubit count per Fat-Tree node scales \emph{linearly} with its height. The router count follows $\sum_{i=0}^{n - 1} (n - i)2^i = 2N - 2 - n$, only doubling that of BB QRAM. The following sections demonstrate that Fat-Tree QRAM significantly improves parallel query latency over sequential BB QRAM queries, with minimal qubit and connectivity overhead.

 \begin{figure*}[t]
         \centering
         \includegraphics[width=0.95\textwidth]{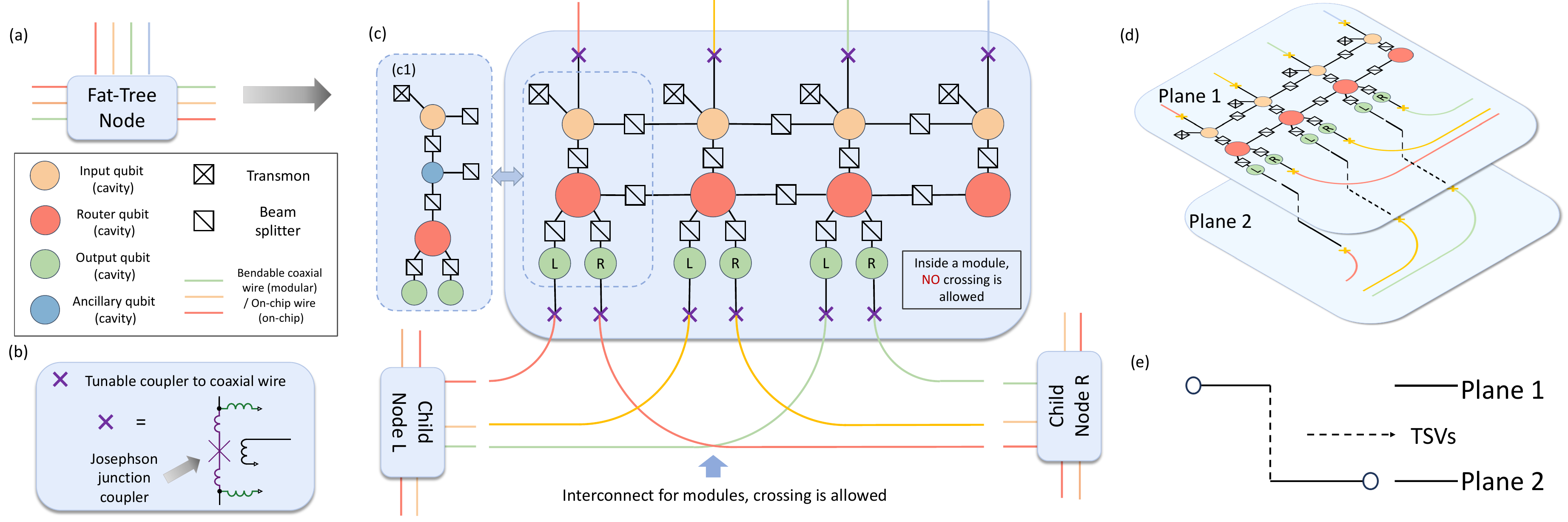}
         \caption{Internal structure of a Fat-Tree node. (a) An example node $(i=1,j=0)$ in a capacity-32 Fat-Tree QRAM, containing 4 routers, 4 incoming wires from the top, and two sets of 3 outgoing wires to its left and right children. (b) A tunable coupler to coaxial wire for inter-node connections in modular design (Sec.~\ref{subsubsec:modular}). (c) The internal structure of a multiplexed router based on superconducting cavities. Transmon-attached input qubit and router qubit ensure native (cavity-controlled) \texttt{CSWAP} gate implementation \cite{weiss2024quantum}. Beam splitters between routers provide intra-node connectivity for local swapping operations. (Sec.~\ref{subsec:operation}) Inset (c1) is an alternative implementation of the router unit enclosed in dashed box that uses more cavities to reduce the connectivity requirement. (d) On-chip two-layer architecture for Fat-Tree QRAM. (Sec.~\ref{subsubsec:onchip}) (e) Sectional view of on-chip design in (d). Inter-plane connection is achieved by the Through-Substrate-Vias (TSVs) technology.}
         \label{fig:hardware}
\end{figure*}

\begin{figure}[t]
         \centering
         \includegraphics[width=0.485\textwidth]{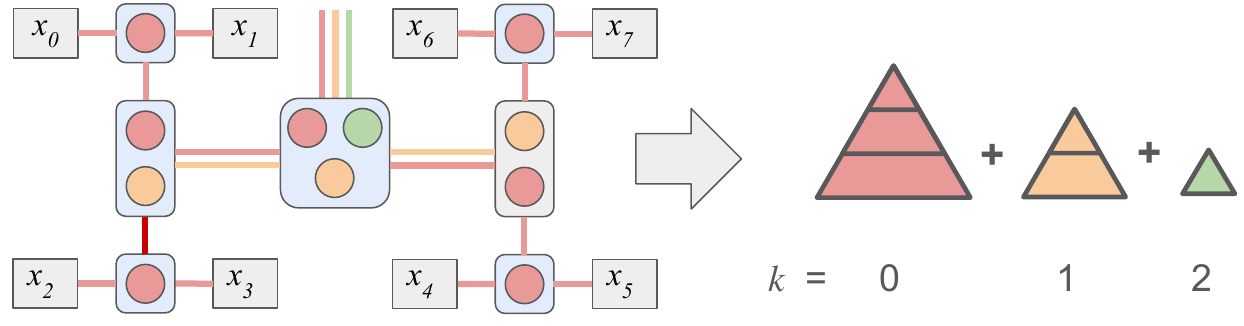}
         \caption{An alternative conceptual interpretation of Fat-Tree QRAM as a composition of multiple BB QRAMs of variable size (Sec.~\ref{subsec:operation}).}
         \label{fig:small}
\end{figure}

\subsection{Implementing Fat-Tree QRAM Nodes}
\label{subsec:nodes}
When choosing a hardware platform for implementing QRAM, we need to consider several essential requirements: (i) efficient encoding of quantum router  (e.g., $\ket{W}, \ket{0}, \ket{1}$), (ii) parallel routing operations (e.g., \texttt{SWAP} and \texttt{CSWAP} gates), (iii) parallel writing of classical data into the state of the bus. 

In this section, we consider implementing a multiplexed Fat-Tree node $(i,j)$ based on superconducting cavities. With rapid advances in superconducting devices, the key elements required in Fat-Tree QRAM have already been proposed, e.g., \texttt{CSWAP} operations between superconducting cavities \cite{gao2019entanglement, xue2023hybrid} and transmon devices \cite{miao2023implementation}. These advances in hardware enable two possible hardware implementations of QRAM, using well-established encodings of qubits (i.e., transmons and cavities), beam-splitters, wires, and tunable couplers. 

For Fat-Tree QRAM, it is important to also consider the intra- and inter-node connectivity caused by the extra routers. For example, in Fig.~\ref{fig:hardware}(c), we provide the internal structure of node $(1,j)$ in a capacity-$N=32$ Fat-Tree QRAM from Fig.~\ref{fig:fat}. In this node, there are four routers, each with four input wires and two sets of three output wires allocated to both child nodes, denoted as L and R. When multiple routers are positioned in a Fat-Tree node, the two output qubits from each router must be routed towards the two external output interfaces (i.e., L and R directions), resulting in possible wire crossings. However, the connectivity constraint for Fat-Tree QRAM is \emph{not} all-to-all. Instead, we show that a \emph{bi-planar nearest-neighbor} connectivity is sufficient. This important observation leads to the efficient implementations of Fat-Tree QRAM using readily available technologies in superconducting circuits, one of the platforms with the most restrictive connectivity constraints. We illustrate this by introducing modular and on-chip architectures for Fat-Tree nodes.

\subsubsection{Fat-Tree Node: Modular Implementation}
\label{subsubsec:modular}
The modular implementation allows us to manufacture all the nodes as independent modules and link them with superconducting coaxial cables \cite{zhong2021deterministic}. Fig.~\ref{fig:hardware} proposes a possible implementation consisting of two fundamental components: (1) tunable couplers with coaxial wires to provide inter-node connectivity and (2) quantum routers inside the module for executing \texttt{CSWAP} gates. Within the module, routers are arranged side by side, with the last router lacking output qubits and serving as a transient storage for queries, resulting in one fewer output wires compared to inputs. Within each quantum router, qubits are constructed by cavities, featuring a transmon coupled to the input cavity for native \texttt{CSWAP} gates implementation~\cite{weiss2024quantum,chapman2023high}. Additionally, horizontal nearest-neighbour connectivity among routers is implemented by beam splitters, enabling swap gates between adjacent routers within Fat-Tree nodes via manipulation of nearest-neighbor coupled qubits along the line. Since the router qubit having four beam splitters attached may cause hardware manufacturing challenges, we provide an alternative implementation to reduce connectivity requirements by adding one extra cavity in Fig.~\ref{fig:hardware}(c1).

The tunable couplers are aligned to the top and bottom of the chip, coupled with the input and output qubits as ports to inter-node wires. Since the coaxial wires can be twisted to any shape, the crossing of routings inside a node is reduced to the crossing of wires connecting different nodes, with no crossings inside the module, as shown in Fig.~\ref{fig:hardware}.

\subsubsection{Fat-Tree Node: On-chip Implementation}
\label{subsubsec:onchip}
In contrast to modular designs, an on-chip implementation of shared QRAM integrates all components onto a single chip, resulting in a significantly reduced size. This approach offers multiple advantages, including faster cooling, reduced energy dissipation, and enhanced fidelity, at the expense of connectivity constraints~\cite{ganjam2024surpassing}. Instead, qubits and wires must be arranged in a planar layout without overlapping. While achieving this in a single-layer chip poses challenges, we demonstrate that the connectivity graph of a Fat-Tree QRAM can be effectively decomposed into two planar subgraphs. Consequently, a thickness-2 chip, consisting of two edge-disjoint layers, can be implemented, as shown in Fig.~\ref{fig:hardware}(d). Although a single-layer chip is preferable for hardware simplicity, employing two layers may still be feasible, as couplers and their control lines can be attached to the top and bottom of the chip. The connection between the two planes can be facilitated with the TSVs technique introduced in Sec.~\ref{sec:background}.

Fig.~\ref{fig:hardware}(e) depicts a sectional view of the on-chip TSVs implementation with the vertical dashed line between the two planes. A single node, implemented by the routers and couplers discussed in the modular design, resides fully in a single plane. However, only one of its child nodes resides in the same plane, while the other resides in the opposite, as the L/R output qubits of quantum routers direct to the chip's opposite/same plane respectively. This alternating plane configuration ensures that wires do not intersect within any single plane, providing an efficient two-plane decomposition.

\subsection{Operations in Fat-Tree QRAM}
\label{subsec:operation}
With the hardware architecture defined, we now describe the efficient operations in Fat-Tree QRAM to implement quantum queries. In this section, we specify the step-by-step inter- and intra-node operations in Fat-Tree QRAM to realize $O(\log(N))$ quantum queries in $O(\log^2(N))$ circuit depth. 

To appreciate the benefits of the proposed architecture, it is useful to consider an alternative interpretation of Fat-Tree QRAM. As shown in Fig.~\ref{fig:small}, if we look at the routers of different colors in isolation, a Fat-Tree QRAM is equivalent to a composition of multiple ``sub-component QRAMs'' of varying sizes (each with address width ranging from $n$ to 1, which we denote by parameter $k$). Fig.~\ref{fig:small} shows the largest QRAM corresponding to $k=n-1$, while the smallest is parameterized by $k=0$. In the Fat-Tree, we index routers belonging to QRAM $k$ as $(i,j,k)$, where each node $(i,j)$ in Fat-Tree incorporates exactly one router from QRAM $k$ if $i \leq k$, and no router if $i > k$. 

Each quantum query involves routing and swapping among the sub-component QRAMs: starting from the smallest-size QRAM at the initial step of the query, we transition to the QRAM larger by one size every time a level of routers is loaded. This can be realized by introducing a \emph{swap step} between consecutive \emph{gate steps} within the original QRAM circuit, as illustrated in Fig.~\ref{fig:fullpip}. 

In Fig.~\ref{fig:fullpip}, the gate steps, highlighted in a white background, load/unload a single layer in the QRAM through a series of \texttt{CSWAP}s, the same as those in a conventional BB QRAM address loading circuit, taking 4 circuit layers each. Meanwhile, the newly introduced swap steps, on a gray background, perform two distinct functions: (i) During the address loading (unloading) stage, the entire query is swapped to a larger (smaller) sub-component QRAM, taking a single circuit layer. All swap gates can be performed in parallel within a Fat-Tree node. (ii) Data retrieval operations are executed during a swap step as well. Data retrieval, similar to in BB QRAM, takes 1 circuit layer for the classically controlled gates. However, in practice, each circuit layer in a swap step takes only a fraction of the time as other circuit layers since intra-node and classical gates are much faster.

For Fat-Tree QRAM, additional time for swapping classical memory might be required for executing multiple distinct queries. We quantify the time budget for classical memory swap without causing query slowdown in Sec.~\ref{sec:results}.

\subsubsection{Pipelining Details}

\begin{figure*}[t]
         \centering
         \includegraphics[width=0.75\textwidth]{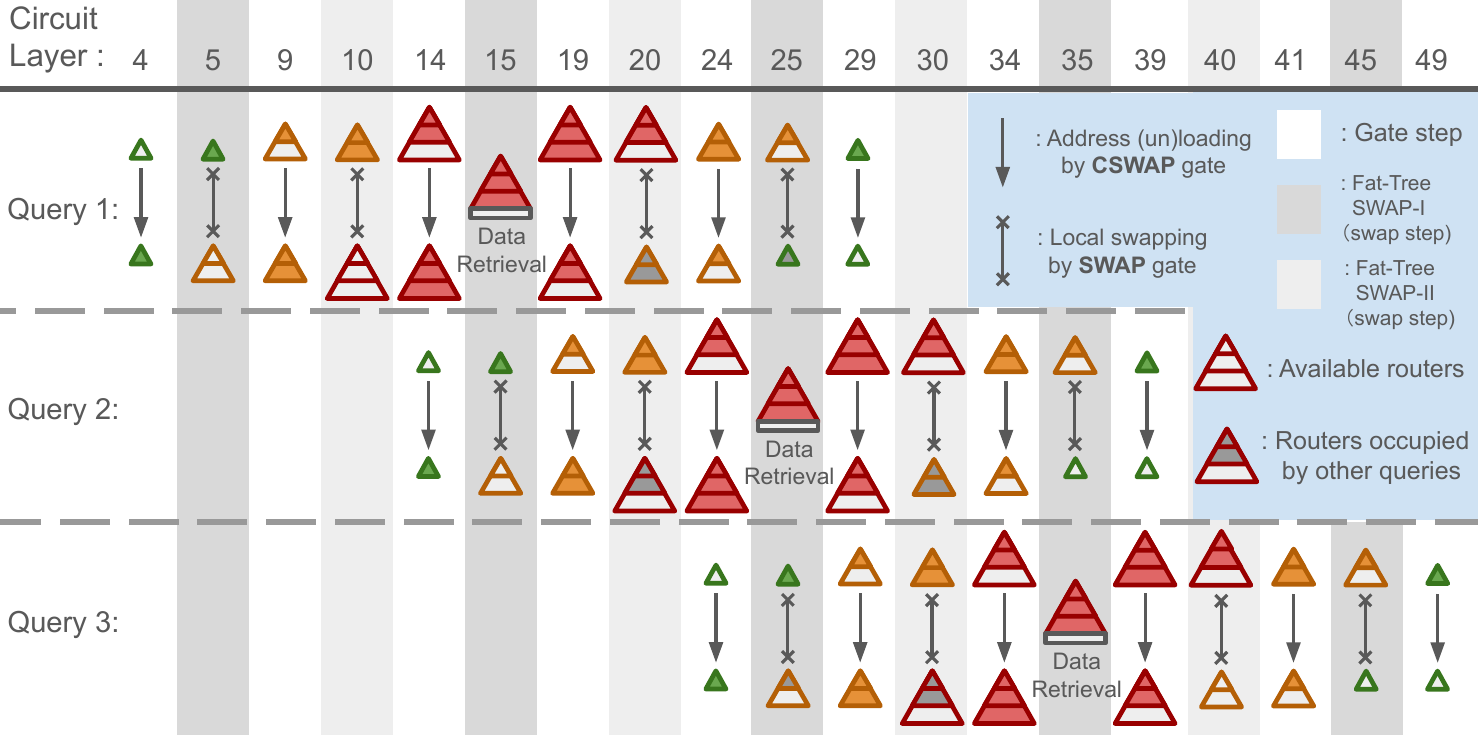}
         \caption{Pipeline schedule of a capacity-8 Fat-Tree QRAM running 3 concurrent queries. Colors indicate which conceptual QRAM $k$ in Fig.~\ref{fig:small} is being used at the hardware level. No conflicting colors in the same layer ensures no concurrent access to the same quantum routers. The latency overhead for each single query of Fat-Tree QRAM compared to the original BB QRAM (query latency 29:25 in the case $n=3$) comes from additional swap steps except the one coinciding with data retrieval, as it is included in BB QRAM's latency.}
         \label{fig:fullpip}
\end{figure*}

Fat-Tree QRAM introduces \emph{query-level pipelining}, unlike the bit-level pipelining of BB QRAMs, enabling greater parallelism without additional resources. We illustrate this pipelining process by integrating circuit gates with additional swap operations in Fig.~\ref{fig:fullpip}.  

QRAM swapping, while seemingly complex, is efficiently executed via \emph{local swapping}, where each node independently swaps internal qubits. Since quantum queries involve all QRAM branches in superposition, local swapping provides a constant-depth solution by performing swaps within each node $(i,j)$ rather than requiring inter-node communication. Specifically, swapping QRAMs $k$ and $k+1$ involves each node $(i,j)$ swapping router $(i, j, k)$'s input and router qubits with those in $(i,j,k+1)$.  

Fig.~\ref{fig:fullpip} defines two local swapping types:  
- \texttt{SWAP-I}: Even-indexed routers $(i,j,k) \Leftrightarrow (i,j,k+1)$ for even $k$.  
- \texttt{SWAP-II}: Odd-indexed routers $(i,j,k) \Leftrightarrow (i,j,k+1)$ for odd $k$.  

The smallest QRAM ($k=0$) does not undergo Type-II swaps, and the largest QRAM ($k=n-1$) swaps only once, depending on the parity of $n$. Alternating \texttt{SWAP-I} and \texttt{SWAP-II} enables seamless query movement across QRAM sizes, facilitating both loading and unloading. Additionally, local swapping maintains sequential qubit allocation with only nearest-neighbor connectivity, simplifying hardware design (Fig.~\ref{fig:hardware}(c)).  

Local swapping requires only a single circuit layer and can run concurrently with classical data retrieval(note that only one type of local swapping will be associated with data retrieval, with the type depending on the parity of $n$). The pipeline interval spans 10 circuit layers, structured as: gate step (4) + \texttt{SWAP-I} (1) + gate step (4) + \texttt{SWAP-II} (1), ensuring efficient scheduling regardless of $n$’s parity.  

We summarize the $\log(N)$-pipelined query procedure in Alg.~\ref{alg:query} and visualize it in Fig.~\ref{fig:fullpip}. A detailed breakdown with elementary operations appears in Appendix Fig.~\ref{fig:detailbb}. The key conceptual steps are as follows:

\begin{enumerate}[label= (\alph*)]
    \item Start a new query and execute one gate step of address loading/unloading for all existing queries.
    \item Apply Type-I swap step, together with classical data retrieval for the leaves of Fat-Tree QRAM if address loading is finished for one of the ongoing queries.
    \item Apply one gate step of address loading/unloading for existing queries. 
    \item Apply swap step Type-II, with data retrieval if needed.
    \item Repeat (a-d) until all the query requests are served.
\end{enumerate}

\begin{algorithm}[h]
\caption{Pipeline $\log(N)$ Quantum Queries with size-$N$ Fat-Tree Shared QRAM.}\label{alg:query}
\begin{algorithmic}[1]

\State \textbf{Require: } $\ket{\psi_{A}^{j}} = \sum^{N-1}_{i=0} \alpha_i^j\ket{i}$
\Comment{address of the $j$th query}
\State \textbf{Require: } $0 \leq j \leq n-1$ 
\Comment{query index}
\State \textbf{Require: } $ n=log(N) \geq 1$
\Comment {QRAM height/total queries}

\State \textbf{Ensure: } $\ket{\psi_{AB}^{j}} = \sum_{i=0}^{N-1}\alpha_i^j\ket{i}_A\ket{x^j_i}_B$
\Comment{$j$th query}

\For{$t = 1, 2, ...$ until queries are finished}
\If{$t$ is odd}
    \If{$t \equiv 1\mod 4$}
        \State Start next query $\ket{\psi_{A}^{(t - 1) / 4}}$  
    \EndIf
    \State Load/Unload Layer (Alg. \ref{alg:load} \& \ref{alg:unload}) $\forall$ existing queries
\Else  
    \If{$t \equiv 2\mod 4$}
        \State \texttt{SWAP-I}: $(i,j,k) \Leftrightarrow(i,j,k+1)$ $\forall$ even $k$
        \If{$n$ is odd}
            \State \texttt{CLASSICAL-GATES}
            \State {} \Comment{Data retrieval for fully loaded query}
        \EndIf
    \Else
        \State \texttt{SWAP-II}: $(i,j,k) \Leftrightarrow(i,j,k+1)$ $\forall$ odd $k$
        \If{$n$ is even}
            \State \texttt{CLASSICAL-GATES} 
            \State {} \Comment{Data retrieval for fully loaded query}
        \EndIf
    \EndIf
\EndIf
\EndFor
\end{algorithmic}
\end{algorithm}

\section{Scheduling Quantum Queries}
\label{sec:schedule}

The previous section examined the architectural design and hardware implementation of Fat-Tree QRAM. Optimizing query performance, however, also depends on efficient QPU-QRAM collaboration at the compiler level, particularly in scheduling query requests. This section introduces a latency-optimal scheduling algorithm for Fat-Tree QRAM and explores its full utilization by analyzing intrinsic quantum algorithm structures.

\subsection{Increasing Utilization of a Shared QRAM}

\begin{figure*}[t]
     \centering
         \centering
         \includegraphics[width=0.7\textwidth, trim=0 1.5em 0 0]{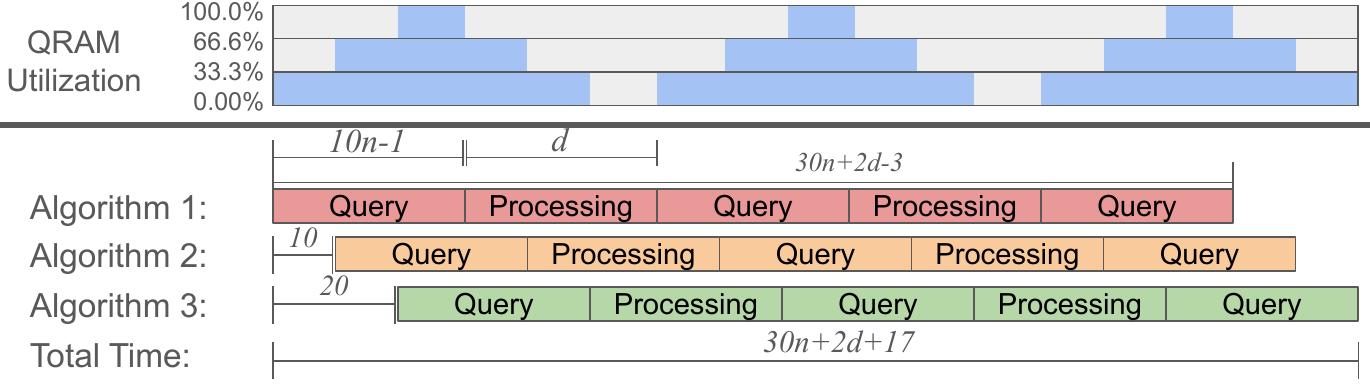}
         \caption{Algorithm execution and query scheduling diagram with Fat-Tree QRAM. Every single query requires $10n-1$ circuit layers to finish for address width $n$, followed by $d$ circuit layers of QPU processing before the next query. In this example, QRAM is underutilized; that is, additional queries can be pipelined.}
         \label{fig:algopip}
\end{figure*}

Fig.~\ref{fig:fullpip} illustrates that $\log(N)$ queries can be efficiently pipelined in $O(\log(N))$ circuit layers if executed consecutively. However, real algorithms may not issue queries uniformly, as seen in Fig.~\ref{fig:algopip}, which incorporates processing stages occupying $d$ circuit layers between queries. If queries occur at regular intervals of depth $d$, QRAM utilization will sometimes be below 1.

State-of-the-art QRAMs serve requests sequentially, yielding binary utilization (0 or 1). In contrast, a capacity-$N$ Fat-Tree QRAM pipelines $\log(N)$ queries, allowing utilization to vary between 0 and 1, enabling additional queries during QPU processing intervals. This suggests Fat-Tree QRAM can accommodate $\log(N)+d$ distributed QPUs rather than just $\log(N)$. Understanding algorithmic structures is key to maximizing QRAM utilization, especially in shared QRAM architectures interacting with multiple QPUs, further explored in Sec.~\ref{sec:results}.

\subsection{Offline and Online Query Scheduling}

The above discussion considers offline scheduling, where query intervals are predetermined. In practice, shared QRAM must handle online query requests, making scheduling more complex as QRAM lacks prior knowledge of QPU activity, and queries arrive at random intervals.

Using a greedy exchange proof, we demonstrate that First-In-First-Out (FIFO) scheduling minimizes total query latency for both offline and online cases (proof in Sec.~\ref{subsec:greedyproof}). Assume an optimal schedule deviating from FIFO, where a later-requested query is processed first. Swapping these queries to align with the request order does not worsen latency. Repeatedly applying this swap to all out-of-order query pairs transforms the schedule into FIFO while maintaining non-increasing latency. Thus, FIFO scheduling is optimal, ensuring minimal total latency.

\section{Evaluation Methodology}
\label{sec:eval}

\subsection{Baseline Architectures}

In this paper, we analyze the performance of Fat-Tree QRAM, Distributed Fat-Tree QRAM (D-Fat-Tree), BB QRAM\cite{giovannetti2008architectures}, Distributed Bucket-Brigade QRAM (D-BB), and Virtual QRAM (Virtual) \cite{xu2023systems}. For Fat-Tree and BB, we assume a single QRAM of capacity $N$ is used as a shared memory.  For D-Fat-Tree, we assume $\log(N)$ distributed BB QRAMs of capacity $N$ are used. For D-BB, we assume $\log(N)$ distributed BB QRAMs of capacity $N$ are used. For Virtual QRAM, we create $\log(N)$ virtual QRAMs, each using $O(N/\log(N))$ qubits to access a large address space ($N$) at the expense of increased latency. More concretely, it divides the capacity $N$ into $K$ pages with size $M=N/K$ for each page and constructs a QRAM with $O(K\log{(M)})$ query latency and $O(M+\log{(K)})$ qubit counts. For a fair comparison, the Virtual baseline uses the same total number of qubits as Fat-Tree. In the subsequent results, the baselines are organized into two groups: the first group, comprising Fat-tree, BB, and Virtual, utilizes $(O(N))$ qubits, while the second group, including D-Fat-Tree and D-BB, requires $(O(N\log(N)))$ qubits—an asymptotically greater quantity than the first group.

\subsection{Quantitative Performance Metrics for QRAM}
To quantify the performance of QRAM, especially under a parallel query setting, we define several important metrics: (i) query parallelism, (ii) max query rate, (iii) QRAM bandwidth. \emph{Query parallelism} is the maximum number of parallel queries a QRAM can execute simultaneously. We say a QRAM is under utilized if it is executing queries fewer than the allowed query parallelism. \emph{Max query rate} (in units of queries per sec) is defined as the maximum number of queries completed per unit time. In our pipelined Fat-Tree QRAM, this is calculated by inverting the amortized time of a single query. Finally, we define QRAM \emph{bandwidth} (in units of qubits per second) as the rate at which data are queried and written into bus qubits, which can be calculated by the product of max query rate and bus width (i.e., number of data qubits returned per query). We focus on results with bus width $= 1$ in Sec.~\ref{sec:results}.

To assess QRAM performance, it is useful to isolate hardware capabilities by using device-independent metrics like \emph{circuit layers}. A circuit layer is one logical step where all quantum gates within the same layer are executed in parallel. It can be further combined with the device clock speed, e.g., Circuit Layer Operations Per Second (CLOPS) \cite{amico2023defining}, to quantify the actual hardware performance.

\subsection{Benchmarks and Applications}
A shared QRAM can facilitate running multiple algorithms in parallel or running a parallel quantum algorithm. We benchmark the benefits of Fat-Tree QRAM for both synthetic and real-world algorithms:

\emph{Synthetic algorithms:} We define a family of circuits, each with alternating processing (for time $d$) and query (for time $t_1$). We test ratios of $d/t_0$ ranging from 0 to 2. Section~\ref{sec:utibench} presents benchmarking results of synthetic algorithms, each repeating querying and processing 10 times with capacity $N = 1024$. These algorithms enable the comparison of QRAM utilization and overall algorithm depth between different QRAM architectures.

\emph{Parallel Grover’s search:} Grover's algorithm can be  applied in parallel to $p$ segments of the database~\cite{zalka1999grover}, where each segment is queried $O\left(\sqrt{N/p}\right)$ times.

\emph{Parallel $k$-Sum:} By implementing $p$-parallelized queries to create modified states for the quantum walk, the parallel $k$-Sum algorithm improves the query complexity from $O(N^{k/k+1})$ to $O({(N/p)}^{k/k+1})$.

\emph{Parallel Hamiltonian Simulation}: Some structured Hamiltonian simulations can be implemented by parallel quantum walks \cite{zhang2024parallel}.

\emph{Parallel Quantum Signal Processing (QSP)}: By factoring degree-$d$ polynomials to the product of $p$ smaller polynomials of degree $O(d/p)$ \cite{martyn2024parallel}, the parallel QSP improves the query complexity from $O(d)$ to $O(d/p)$.

\section{Results}
\label{sec:results}

\subsection{Resource Estimation and Comparison}
Table \ref{tab:resource} presents a comprehensive comparison of different shared QRAM implementations with various parameters, including qubit count, query parallelism, and query latency. All numbers in the table are precise counts for a QRAM with capacity $N$. Baseline BB is a sequential query architecture suffering from $O(\log^2(N))$ overhead in query latency.  Comparing Fat-Tree QRAM to the same-qubit-count baseline, Virtual, the overall query latency for parallel queries is asymptotically slower than Fat-Tree QRAM, due to the large single query latency in Virtual, namely $O(\log^2(N))$ for Virtual QRAM vs $O(\log(N))$ for BB. In particular, the Virtual architecture decomposes the total address space $N$ into $K= \log(N)/2$ pages, where each page has size $M= N/\log(N)$ and requires a native Multi-Control-X (MCX) gate to implement. Baseline D-BB has low query latency while requiring asymptotically more resources. It is asymptotically slower in multiple-query tasks compared to the same-qubit-count baseline D-Fat-Tree QRAM. Consequently, under the same resource constraints, the Fat-Tree QRAM asymptotically outperforms state-of-the-art QRAM architectures regarding overall query latency for parallel query requests. All our resource estimation in Table \ref{tab:resource} based on realistic hardware parameters under the recent improvement in the implementation of native \texttt{CSWAP} gates, offering a $\tau=1\mu s$ gate time~\cite{weiss2024quantum}, or equivalently clock speed $1/\tau= 10^6$ CLOPS~\cite{amico2023defining}. The intra-node SWAP gate is even faster with gate time $T_{\textrm{SWAP}}=125 ns$~\cite{liu2024quantum,weiss2024quantum}.

\begin{table*}[t]
\small
    \centering
    % \begin{tabular}{ p{3cm}||p{3.8cm}|p{1.8cm}|p{1.8cm}|p{4.7cm}}
    \begin{tabular}{ c||c|c|c|c|c}
     \hline
     \hline
       & Fat-Tree & D-Fat-Tree & BB \cite{giovannetti2008architectures} & D-BB & Virtual \cite{xu2023systems} \\
     \hline
     \makecell[l]{Qubits} & $16N$ & $16N\log(N)$ & $8N$  & $8N\log(N)$ & $16N$  \\
     \hline
     Query parallelism $m$ & $\log(N)$ & $\log^2(N)$ & $1$ & $\log(N)$ & $\log(N)$\\
     \hline
     \makecell[l]{Query latency for \\ single query $t_1$} & \makecell[l]{$8.25\log(N)$ \\ $- 0.125$} & \makecell[l]{$8.25\log(N)$ \\ $- 0.125$} & $8\log(N) + 0.125$ & \makecell[l]{$8\log(N) + 0.125$} & \makecell[l]{$4\log^2(N) + 4.0625\log(N)$\\$- 4\log(N)\log(\log(N))$} \\
     \hline
     \makecell[l]{Query latency for \\$\log(N)$-parallel \\ queries $t_{\log(N)}$ } & \makecell[l]{$16.5\log(N)$ \\ $- 8.375$} & \makecell[l]{$16.5 - \frac{8.375}{\log(N)} \;^*$} & \makecell[l]{$8\log^2(N)$\\$+ 0.125\log(N)$} & \makecell[l]{$8\log(N) + 0.125$} & \makecell[l]{$4\log^2(N) + 4.0625\log(N)$\\$- 4\log(N)\log(\log(N))$} \\
     \hline
     \makecell[l]{Amortized Single \\ Query Latency} & $8.25$ & $\frac{8.25}{\log(N)}$ & $8\log(N) + 0.125$ & $8 + \frac{0.125}{\log(N)}$ & \makecell[l]{$4\log(N) + 4.0625$\\$- 4\log(\log(N))$} \\
     \hline
    \end{tabular}
    \caption{Space (i.e., qubit number) and time (i.e., query latency) resource comparison across different shared QRAM models with classical memory size $N$. Latency is calculated with intra-node and classical gates, taking only an eighth of the time as a standard circuit layer. Compared to BB QRAM, Fat-Tree QRAM achieves an asymptotic reduction in query latencies for $\log(N)$ parallel queries at the cost of constant overhead (i.e., doubling) in qubits. Note that $t_{\log(N)}$ for D-Fat-Tree is the amortized time for $\log(N)$ queries since D-Fat-Tree has a higher parallelism than $\log(N)$ queries (i.e. $\log(N)$ queries is insufficient to fully utilize D-Fat-Tree).}
    \label{tab:resource}
\end{table*}

\begin{table*}[t]
    \small
    \centering
    \begin{tabular}{c||c|c|c|c|c}
     \hline
     \hline
       & Fat-Tree & D-Fat-Tree & BB \cite{giovannetti2008architectures} & D-BB & Virtual \cite{xu2023systems}  \\
     \hline
     \makecell[l]{QRAM bandwidth \\ (qubit/sec)} & $1.21 \times 10^5$ & \makecell[l]{$1.21\log(N) \times 10^5$} & $\frac{1.25\times 10^5}{\log(N)} + 8 \times 10^6$ & \makecell[l]{$\frac{10^6 \log(N)}{8\log(N) + 0.125}$} & $\frac{10^6}{4\log(N) + 4.0625 - 4\log(\log(N))}$ \\
     \hline
     \makecell[l]{Space-time Volume \\ per query } & $132N$ & $132N$ & $64N \log(N) + N$ & 64N $\log(N) + N$ & \makecell[l]{$64N\log(N) + 65N $\\$- 64N\log(\log(N))$} \\
     \hline
     \makecell[l]{Time budget for classical \\ memory swap ($\mu s$) } & $8.25$ & $8.25$ & $8\log(N) + 0.125$ & $8\log(N) + 0.125$ & \makecell[l]{$4\log^2(N) + 4.0625\log(N)$\\$- 4\log(\log(N))$} \\
     \hline
    \end{tabular}
    \caption{Bandwidth, memory access rate, and space-time volume comparison across different shared QRAM models with classical memory size $N$. The \texttt{CSWAP} gate time is estimated at $1 \mu s$ \cite{weiss2024quantum}, which leads to QRAM clock speed at $1 \times 10^6$ circuit layer operations per second (CLOPS) \cite{amico2023defining}. Fat-Tree QRAM achieves a high bandwidth and memory access rate that is independent of the memory size $N$, and requires asymptotically less space-time volume than other QRAM models.}
    \label{tab:bandwidth}
\end{table*}

\subsection{QRAM Bandwidth}

\begin{figure}[t]
         \centering
         \includegraphics[width=0.96\linewidth]{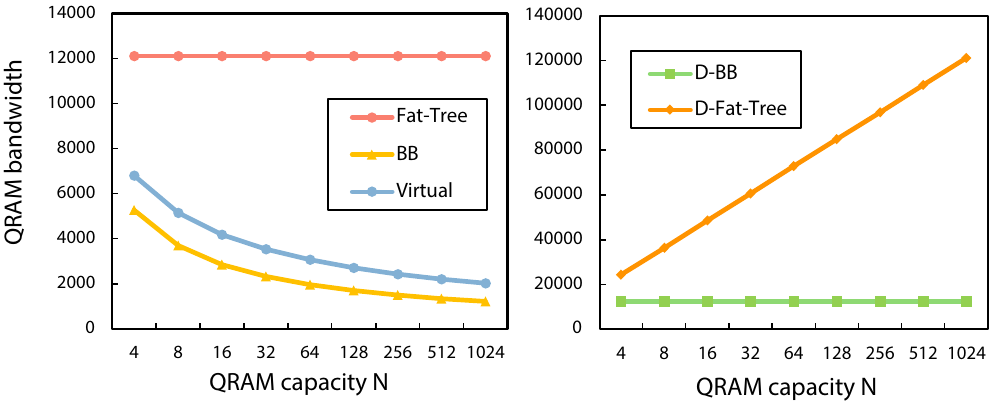}
         \caption{Bandwidth comparison for different QRAM architectures. Fat-Tree achieves a capacity-independent constant bandwidth.}
         \label{fig:bandwidth}
\end{figure}

Table \ref{tab:bandwidth} includes a comparison of QRAM \emph{bandwidth} for all the QRAM architectures. Recall that in Sec.~\ref{sec:eval}, we define bandwidth as the rate at which data qubits can be provided to the QPUs. Fig.~\ref{fig:bandwidth} provides a fine-grained, accurate scaling of QRAM bandwidth and space-time comparison under the same gate parameters in resource estimation.

As shown in Table~\ref{tab:bandwidth}, Fat-Tree QRAM achieves a \emph{constant} bandwidth (i.e., independent of the QRAM size $N$), giving it an asymptotic advantage compared to all other architectures that use the same resources. Though the bandwidth of Baseline D-BB is also constant, it achieves constant scaling at the price of $\log(N)$ copies of the hardware. The bandwidth is also related with \emph{quantum volume per query}, defined as the amortized $\textrm{qubit} \cdot \textrm{circuit depth}$ per query, quantifying the cost of implementing a single query. Fat-Tree QRAM, under the same resource constraints, asymptotically outperforms BB and Virtual QRAM.

Another related metric is \emph{memory access rate} which quantifies the rate at which classical data is read by the QRAM hardware. While this rate is consistent with the bus qubit throughput quantified by bandwidth, the duty rate of shared QRAMs can be simply calculated by $(\textrm{bandwidth} \cdot N)$. 

Finally, we estimate the time budget for classical memory swap using the time interval between two separate queries' data retrieval. While classical memory changes were neglected in previous analyses, large memory shifts can introduce delays in practice if the swap time budget is insufficient. We observed that different architectures pose different classical memory challenges: Fat-Tree requires rapid swapping with constant intervals; in contrast, D-BB requires parallel memory swapping, as the classical memory cells are also copied and distributed for D-BB QRAM. 

\subsection{Enabling Parallel Algorithms}
One of the most significant applications for Fat-Tree QRAM is supporting parallel quantum algorithms requiring parallel queries. In particular, we removed all the dependencies on other parameters by setting them as constant values, such as deviation $\epsilon$ and Hamiltonian sparsity $d$. Hence, the resulting asymptotic scaling only depends on problem size $N$ (capacity of QRAM). Compared to Baselines BB and Virtual, Fat-Tree QRAM achieves the following asymptotic reduction in circuit depth: 
\begin{itemize}
    \item Grover's algorithm: $O(\log^2(N)\sqrt{N})$ to $O(\log(N)\sqrt{N})$
    \item $k$-sum algorithm: $O(\log^2(N)(N/\log(N))^{k/k+1})$ to \newline $O(\log(N)(N/\log(N))^{k/k+1})$
    \item Hamiltonian sim.: $O(\log(N)\log(\log((N))+\log^2(N))$ to $O(\log(N)\log(\log((N))+\log(N))$.
    \item Quantum Signal Processing (QSP): $O(poly(d))$ to \\ $O(poly(d) /\log(N))$, where $d$ is the degree of polynomial encoded in the unitary transformed by QSP.
\end{itemize}

\begin{figure}[t]
         \centering
         \includegraphics[width=\linewidth, trim=0 1em 0 0]{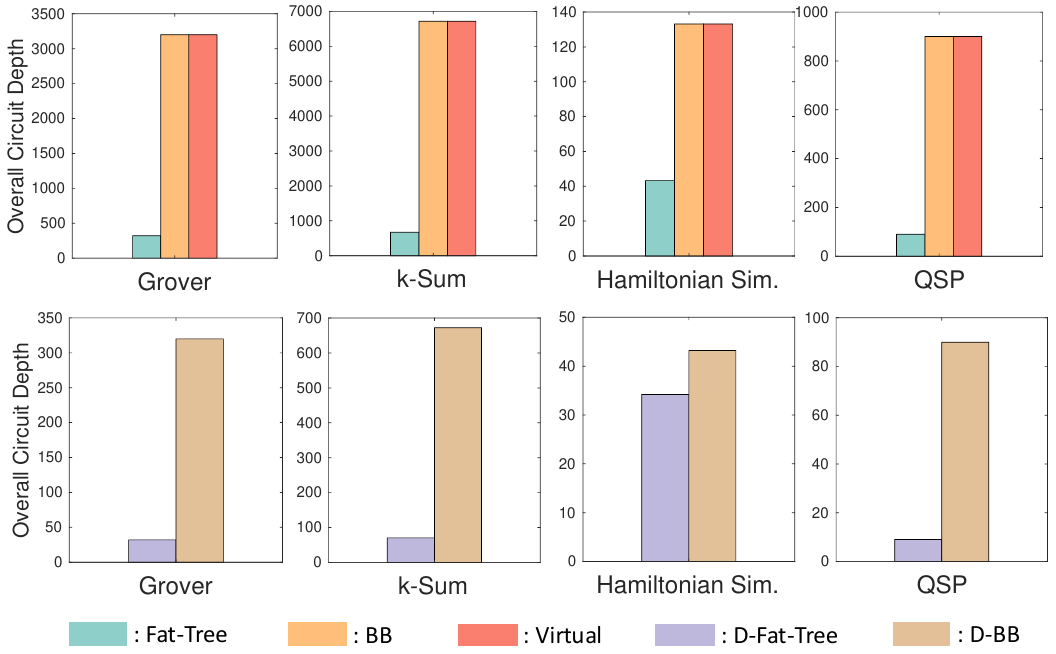}
         \caption{Overall circuit depth for running parallel algorithms, assuming memory size $N=2^{10}$. For QSP, we assume $d=30$ and $poly(d)=d^2$. Fat-Tree QRAM achieves up to a factor of 10 reduction compared to baselines BB and Virtual.}
         \label{fig:parallelalg}
\end{figure}

To put the savings into context, Fig.~\ref{fig:parallelalg} presents concrete examples of overall algorithmic circuit depth reduction (by up to a factor 10) on practical problems with medium-scale memory $N=2^{10}$.

\begin{figure}[t]
         \centering
         \includegraphics[width=0.97\linewidth]{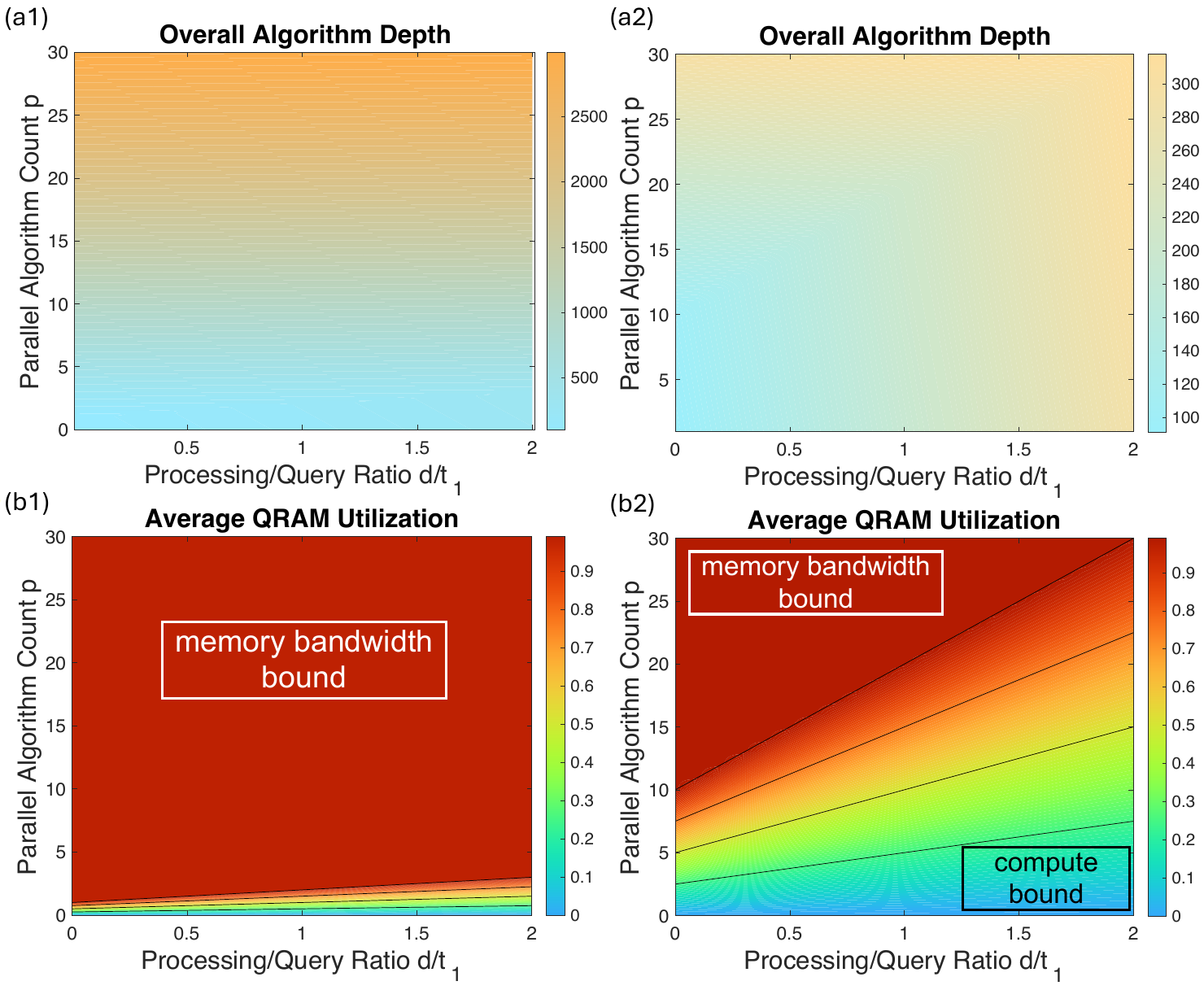}
         \caption{(a1/2) Overall algorithm depth of synthetic algorithm in BB/Fat-Tree QRAM. (b1/2) QRAM utilization of synthetic algorithm in BB/Fat-Tree QRAM. Fat-Tree QRAM balances processing/query ratio and parallel algorithms count, significantly reducing the overall algorithm depth.}
         \label{fig:utilization}
\end{figure}

\subsection{QRAM Hardware Utilization}
\label{sec:utibench}
As discussed in Sec.~\ref{sec:schedule}, to maximize the utilization without queries bottlenecking, an ideal strategy is to allocate a proper number of parallel algorithms to one Fat-Tree QRAM. Fig.~\ref{fig:utilization} presents the result of the synthetic algorithms introduced in Sec.~\ref{sec:eval} and compares the performance of BB with Fat-Tree QRAM on the dependency of processing/query ratio and parallel algorithm count. 
The BB QRAM meets the memory bandwidth bound even with a small increment in parallel algorithm count, resulting in a large overhead in the overall algorithm depth. Our Fat-Tree QRAM, however, is capable of balancing parallel algorithm count and the processing/query ratio, which significantly decreases the depth of synthetic algorithms.

\section{Error Robustness of Fat-Tree QRAM}\label{sec:qec}
In this section, we show that Fat-Tree QRAM maintains BB QRAM's error resilience \cite{hann, xu2023systems} and requires fewer quantum resources for error correction to reach a desired circuit fidelity. Moreover, its capacity for additional queries can enhance query fidelity through virtual distillation \cite{huggins2021virtual} and support error correction. Thus, Fat-Tree QRAM achieves high query fidelity by combining error robustness and mitigation strategies, while remaining compatible with low-overhead QEC techniques for fault-tolerant quantum computing.

\subsection{Noise Resilience of Fat-Tree QRAM}
\label{subsec:noise}

We show that Fat-Tree QRAM maintains the same intrinsic error-resilience and logarithmic infidelity scaling as BB QRAM \cite{hann,xu2023systems}. Following~\cite{hann}, we define \textit{query fidelity} for a single query $\ket{\psi_{in}}$ as $F = \braket{\psi_{out}|\rho_{out}|\psi_{out}}$
where $\rho_{out}$ is the actual output density matrix under noisy channel, and $\psi_{out}$ is the ideal expected output state. For the noise model, we assume that each qubit is subjected to a generic error channel when a gate is applied:
$\mathcal{E}(\rho) = (1-\epsilon) \rho + \epsilon K\rho K^\dag$, where $\epsilon$ is the error probability and  $K$ denotes the error Kraus operator. We show that in the presence of this error channel, the Fat-Tree QRAM has the asymptotic lower bound in query fidelity $F \ge 1 - 2 \cdot \log^2{N}\cdot (\epsilon_0+\epsilon_1+\epsilon_2)$ where $\epsilon_0,\epsilon_1, \epsilon_2$ correspond to the three types of gates introduced in Sec.~\ref{sec:arch}: a total of $\log^2(N)$ number of (intra-node) \texttt{CSWAP} gates each with error rate $\epsilon_0$,  $\log^2(N)$ (inter-node) \texttt{SWAP} gates with error rate $\epsilon_1$, and $\log^2(N)$ local (intra-node) \texttt{SWAP} gates with error rate $\epsilon_2$. Thus, the total fidelity is $F \ge (2 \cdot \prod_i(1-\epsilon_i)^{G_i}-1)^2 \ge 1 - 2 \cdot \log^2{N}\cdot (\epsilon_0+\epsilon_1+\epsilon_2)$.
Compared to BB QRAM's infidelity lower bound of $F \ge 1 -  2\cdot\log^2{N}\cdot(\epsilon_0+\epsilon_1)$ \cite{hann}, Fat-Tree QRAM achieves parallelism with only a moderate decrease in fidelity. Furthermore, Fat-Tree QRAM is compatible with the error robust analysis in \cite{mehta2024analysis}, where this error resilience in QRAM is extended to more generic error models, including initialization errors, spatially correlated errors, and coherent errors.

Fig.~\ref{fig:qecinf} provides a query infidelity comparison between Fat-Tree and BB, where the error rates for the three types of gates are set to experimentally realistic values: $\epsilon_0=0.002, \epsilon_1=0.002, \epsilon_2=0.001$~\cite{weiss2024quantum,zhong2021deterministic,niu2023low}. The infidelity scaling of Fat-Tree QRAM is only a constant factor (0.25x) worse compared to BB, due to intra-node \texttt{SWAP} gates implemented using beam-splitters, which is fast and high fidelity compared to the other two types of gates described above \cite{weiss2024quantum, chapman2023high}. As hardware continues to improve, QRAM of larger capacity will become practical. This is illustrated in Table \ref{tab:error_rates}, for different baseline error rates $\epsilon_0$ with realistic parameters from \cite{weiss2024quantum}.

\begin{table}[t]
\small
\centering
\begin{tabular}{ p{1.5cm}||p{1.5cm}|p{1.5cm}|p{2.2cm}}
\hline
Capacity $N$ & $\epsilon_0=10^{-3}$ & $\epsilon_0=10^{-4}$ & $\epsilon_0=10^{-5}$ (with post-selection) \\ \hline
8 & 0.045 & 0.0045 & 0.00045 \\ \hline
16 & 0.08 & 0.008 & 0.0008 \\ \hline
32 & 0.125 & 0.0125 & 0.00125 \\ \hline
64 & 0.18 & 0.018 & 0.0018 \\ \hline
\end{tabular}
\caption{Query infidelity of QRAM with capacity $N$, given different input gate error rates $\epsilon_0$ from \cite{weiss2024quantum}. The desirable scaling comes from QRAM's intrinsic noise resilience.}
\label{tab:error_rates}
\end{table}

\subsection{Virtual Distillation using Fat-Tree QRAM}
\label{subsec:vd}

In this section, we show how to leverage parallel queries provided by Fat-Tree QRAM to boost query fidelity. Virtual distillation is a quantum error mitigation technique that "distills" a higher-fidelity state from multiple noisy copies~\cite{huggins2021virtual}. Let an $n$-qubit noisy quantum state be $\rho = (1 - \epsilon)\rho_0 + \epsilon \rho_{\text{error}}$ where $\rho_0$ is the ideal (error-free) state, $\rho_{\text{error}}$ is the error component, and $\epsilon$ is the error rate. As for Fat-Tree QRAM, we prepare $k$ identical copies of the noisy QRAM query 
state $\rho$ in parallel, leading to the combined state $\rho^{\otimes k}$. The objective is to approximate a ``distilled'' state, defined as
$\rho_{\text{distilled}} = \frac{\rho^k}{\operatorname{Tr}(\rho^k)}$. Here, $\rho^k$ represents the $k$-th power of $\rho$.
To measure an observable \( O \) with reduced error, we calculate the expectation value using the distilled state: $\langle O \rangle_{\text{distilled}} = \operatorname{Tr}\left( \frac{\rho^k}{\operatorname{Tr}(\rho^k)} O \right)$. This approach effectively amplifies the ideal component $\rho_0$'s contribution to $\rho$ while suppressing the erroneous content in noisy queries $\rho_{\text{error}}$.

Both using 256 qubits, one Fat-Tree (N=16) and two BB QRAMs (N=16) can perform four parallel queries and two parallel queries, respectively. Under independent stochastic errors, Fat-Tree achieves an exponentially higher fidelity after distillation, shown in table \ref{tab:distill}.

\begin{table}[t]
\small
\centering
\begin{tabular}{ p{5.2cm}||p{1.2cm}|p{0.8cm}}
\hline
 & Fat-Tree & 2 BB \\ \hline
Resource state prepared for distillation & 4 & 2 \\ \hline
Fidelity of single query before distillation & 0.84 & 0.872 \\ \hline
Fidelity after distillation & 0.9994 & 0.984 \\ \hline
\end{tabular}
\caption{Fidelity comparison of capacity-4 Fat-Tree and two BB QRAMs (same-qubit-count baseline) before and after virtual distillation. Details are included in Sec.~\ref{subsec:vd}.}
\label{tab:distill}
\end{table}

In general, Fat-Tree QRAM can group $k$ copies for distillation and still provide $\log(N)/k$ parallel queries, thus presenting a trade-off between query parallelism and fidelity.

\subsection{Error Correction in Fat-Tree QRAM}
\label{subsec:qec}
\subsubsection{Error-corrected query using encoded QRAM}
The intrinsic noise resilience of Fat-Tree QRAM mentioned in Section \ref{subsec:noise} also helps reduce quantum error correction (QEC) overhead in the fault-tolerant era. We assume each qubit is encoded in an $[[m, 1, d]]$ code ($m$ for the number of physical qubits, $d$ for the code distance), along with fault-tolerant  \texttt{SWAP} and \texttt{CSWAP} gates. Notably, although this gate set includes a non-Clifford gate, it can still be implemented using transversal non-Clifford gates while circumventing the constraints of the Eastin–Knill theorem~\cite{eastin2009restrictions}. This is because the limited gates in QRAM circuits do not form a universal set.
For instance, the color code supports a transversal CCZ gate~\cite{kubica2015unfolding}. Beyond transversal gates, alternative methods exist that are more efficient than magic state distillation for implementing non-Clifford operations. One such approach is pieceable fault-tolerant gates~\cite{yoder2016universal}, which perform intermediate error checks during gate execution. For example, [5,1,3] code implements a fault-tolerant Toffoli gate this way. This technique could serve as a strong candidate for implementing fault-tolerant CSWAP gates.

Notably, using different error correction architectures for QRAM and QPU may introduce extra code-switching overhead.  However, only a moderate amount of code-switching is necessary, because only the $n$ input address qubits (and 1 bus qubit) from the QPU need to be converted to the QEC code for QRAM. This is a relatively small fraction of all qubits involved, e.g., there are $N = 2^n$ qubits within the QRAM. For instance, a well-studied code teleportation approach converts between two QEC codes of distance $d_1$ and $d_2$, respectively. The procedure requires the $d_1 * d_2$ ancilla qubits per query and $O(1)$ circuit depth to transfer each of the $n$ address qubits and the bus qubit sequentially~\cite{xu2024constant}. Because Fat-Tree QRAM implements queries in a pipeline fashion, the ancilla qubits can be reused for the parallel queries. A similar claim also applies to state-of-the-art quantum low density parity check (LDPC) codes, with a linear number of ancilla qubits and a linear-depth circuit~\cite{cross2024improved}.

\begin{figure}[t]
         \centering
         \includegraphics[width=\linewidth]{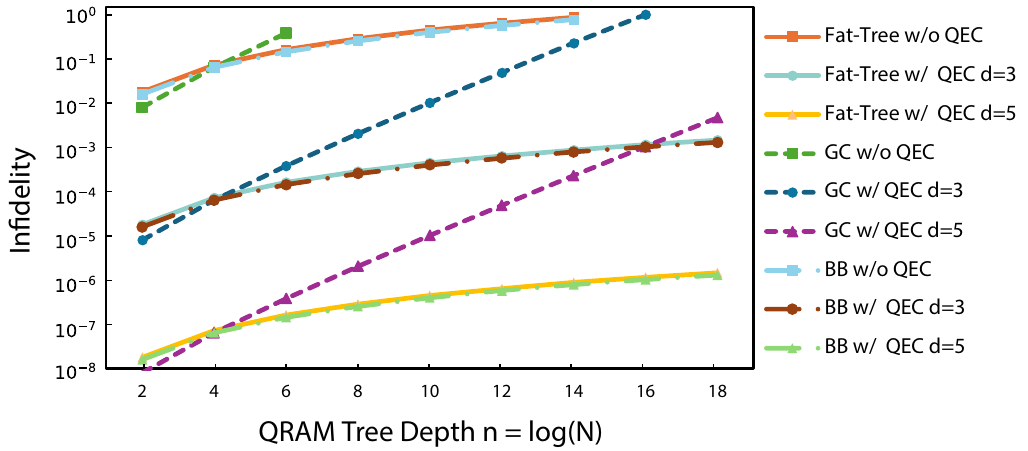}
         \caption{Infidelity of a Fat-Tree QRAM, a BB QRAM, and a generic quantum circuit (GC) as a function of circuit size $N$ (or QRAM tree depth $n=\log(N)$) and QEC code distance $d$, assuming physical gate error rate $\epsilon_0=10^{-3}$. Fat-Tree and BB differ only slightly by a small constant factor, while GC performs exponentially worse than QRAM circuits.}
         \label{fig:qecinf}
\end{figure}

As a result, we can correct any $(d-1)/2$ bit-flip or phase-flip errors at the expense of $O(m\cdot N)$ total qubits. Here we compare with a generic circuit (GC) whose worst-case infidelity scales \emph{linearly} with its circuit size, which is a standard assumption in formal fault tolerance analyses. As shown in Figure \ref{fig:qecinf}, the intrinsic noise resilience protects BB/Fat-Tree QRAM from exponentially decaying fidelity for circuit of growing QRAM tree depth, when compared to a generic circuit. For example, to maintain same infidelity below $5\times 10^{-4}$, QEC with distance 3 allows us to run a GC of tree-depth $n\approx 6$ or a QRAM of tree-depth $n=10$. At the same QEC cost, we can execute a QRAM circuit of larger size than a GC. Conversely, QRAM circuit (of the same size compared to GC) requires a lower QEC code distance to achieve the same fidelity.

When compared to BB QRAM, the infidelity of Fat-Tree QRAM differs by a small constant factor, due to its additional (Clifford) gates. The efficiency of QEC resources for QRAM circuits indicates that Fat-Tree QRAM's robustness could also benefit future fault-tolerant architectures.

\subsubsection{Error-corrected query using noisy QRAM}
Experimental implementations of encoded QRAM can be challenging due to the $O(m\cdot N)$ qubit cost. We further propose a novel scheme that leverages parallel queries on encoded addresses for error protection, without replacing every physical qubit in the QRAM with an encoded logical qubit. Specifically, we assume QRAM qubits are noisy but address/bus qubits are encoded using a QEC code. Finding a QRAM-tailored code is beyond the scope of this work, but we present a resource estimate (in Table~\ref{tab:qecestimation}) for QEC code with parameters $[[m, 1, d]]$ and syndrome extraction circuit of depth $D$.

Fat-Tree enables the encoded address qubits to be routed into the QRAM in parallel. Specifically, due to the fault-tolerant implementation of \texttt{CSWAP}, we can route each of the $m$ physical qubits in an encoded logical address qubit as $m$ pipelined queries. If $m\leq \log(N)$, then $\lfloor \log(N)/m\rfloor$ logical queries can be pipelined. Within each logical query, we can interleave syndrome extraction circuit on qubits from different physical queries reaching the same positions in the QRAM, resulting in a total logical query of depth $O(D\log(N) + m)$. Table~\ref{tab:qecestimation} provides a comparison of this pipelined QEC scheme with an encoded BB QRAM.

\begin{table}[t]
\small
\centering
\begin{tabular}{ p{3.5cm}||p{2cm}|p{1.5cm}}
\hline
 & Fat-Tree &  BB \\ \hline
Physical Qubits & $N$ & $m\cdot N$ \\ \hline
Logical Query Parallelism & $\lfloor \log(N)/m \rfloor$ & 1 \\ \hline
Logical Query Latency & $D\cdot \log(N) + m$ & $D\cdot \log(N)$ \\ \hline
\end{tabular}
\caption{Cost of the error-corrected query (in Big $O$) using (noisy) Fat-Tree QRAM and (encoded) BB QRAM.  We assume an [[$m,1,d$]] ($m\leq\log(N))$ QEC code with transversal \texttt{CSWAP} gate and $D$ syndrome extraction circuit depth. Detailed analysis is included in Sec.~\ref{subsec:qec}. }
\label{tab:qecestimation}
\end{table}

\section{Discussion}\label{sec:discussion}

\subsection{Beyond Superconducting Platforms}
In Section \ref{sec:arch}, we proposed a hybrid cavity-transmon implementation of Fat-tree QRAM, demonstrating that Fat-tree QRAM can be realized even under the stringent connectivity constraints inherent to superconducting architectures. Another promising candidate for this implementation is the trapped-ion platform, which benefits from all-to-all connectivity. By substituting each module in our design with a trapped-ion chip and linking chips through quantum charge-coupled devices (QCCDs), we achieve a scalable Fat-tree QRAM architecture \cite{pino2021demonstration}. Recent advancements in QCCD technology enable multiple operational zones within a single trapped-ion chip, further enhancing the feasibility and practicality of Fat-Tree QRAM deployment at scale \cite{mordini2024multi}.

\subsection{Related Work}
In \cite{paler2020parallelizing}, Paler et al. introduced a "parallel query Bucket Brigade QRAM" based on different query definitions. Their parallel queries refer to classical queries to classical memory, reducing the depth of a single query (quantum query defined in this paper) from $O(N)$ to $O(\log(N))$ by parallelizing Clifford + T gates in the data retrieval stage. This improvement is accounted for and further enhanced by the $O(1)$ data retrieval in Sec.~\ref{sec:background}. However, serving multiple \emph{quantum} queries in a single QRAM remains a highly non-trivial problem and is resolved by our work.

\section{Conclusion}
We presented a blueprint for a shared QRAM architecture based on multiplexed quantum routers in a Fat-Tree structure. It is capable of pipelining multiple quantum queries in parallel, while preserving the space, time, and fidelity scalings as a Bucket-Brigade QRAM. We demonstrate that the hardware architecture can be efficiently implemented on platforms such as superconducting cavities with native \texttt{CSWAP} gates and limited connectivity. Our results suggest Fat-Tree QRAM as a promising architecture for implementing high-bandwidth, noise-resilient quantum queries.
%%
%% The acknowledgments section is defined using the "acks" environment
%% (and NOT an unnumbered section). This ensures the proper
%% identification of the section in the article metadata, and the
%% consistent spelling of the heading.
\begin{acks}
We would like to thank Steven M. Girvin, Liang Jiang, Connor T. Hann, Daniel Weiss, Xuntao Wu, Rohan Mehta, Kevin Gui, Gideon Lee, Allen Zang, and Zhiding Liang for fruitful discussions. This project was supported by the National Science Foundation (under awards CCF-2312754 and CCF-2338063). External interest disclosure: YD is a scientific advisor to, and receives consulting fees from Quantum Circuits, Inc.

\end{acks}

\appendix
\begin{center}
{\Large \bf Supplemental Material for `Fat-Tree QRAM: A High-Bandwidth Shared Quantum Random Access Memory for Parallel Queries'}
\end{center}

\section{Appendix}
\label{sec:appendix}

\subsection{Step-by-step Query Procedure}
\label{subsec:BBdetail}

\begin{figure}[t]
         \centering
         \includegraphics[width=\linewidth]{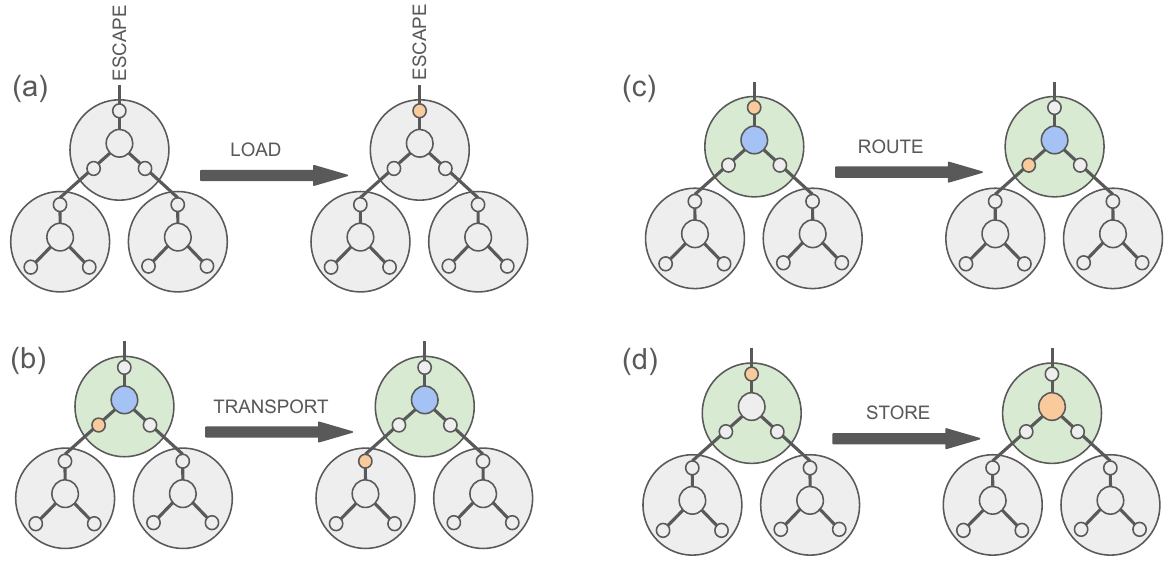}
         \caption{Diagram depicting the effects of the first four fundamental operations: \texttt{LOAD}, \texttt{TRANSPORT}, \texttt{ROUTE}, \texttt{STORE}. The router highlighted in green denotes the router the operation is performed on. The orange qubit depicts where the information is and how the operation moves it while the blue router qubit simply indicates that an address is loaded inside the router.}
         \label{fig:instruction}
\end{figure}

We formally define the elementary QRAM instructions as follows with Figure~\ref{fig:instruction} providing a visual representation of the operations

\begin{enumerate}
    \item \emph{\texttt{LOAD} (L)}: Load operation involves loading a new qubit through the escape to the input qubit of a root router in Fat-Tree QRAM. This operation only happens in the root node of QRAM.
    \item \emph{\texttt{TRANSPORT} (T)}: Transport operation uses a \texttt{SWAP} gate to move a qubit from a router's output qubit to the next level's input qubit
    \item \emph{\texttt{ROUTE} (R)}: Route uses \texttt{CSWAP} gates to route a qubit from the router's input to the outputs according to the state of router qubit. Generally, it is implemented using two \texttt{CSWAP} gates (controlled on the router qubit being 0 and 1), but our computations assume it costs a single circuit layer to simplify complexity and fidelity calculations. Regardless of the implementation, the asymptotic results will remain the same.
    \item \emph{\texttt{STORE} (S)}: Store operation refers to storing an input qubit by swapping it from the quantum router's input qubit to the router qubit. This operation only happens at the highest unloaded layer of a QRAM, and increases the depth of the loaded tree by one.
    \item \emph{\texttt{CLASSICAL-GATES} (CG)}: Performs classically controlled gates to modify the output qubits of the last QRAM level according to values in the classical database.
\end{enumerate}

\algblock{Parallel}{EndParallel}
\begin{algorithm}[t]
\caption{\textsc{Load Layer}}\label{alg:load}
\begin{algorithmic}[1]
    \State \textbf{Require: } $loaded$ as number of address qubits loaded
    \State \textbf{Require: }  $s$ as next address depth to be stored
    \State \textbf{Require: } $k$ as the current QRAM copy being used
    \State \textbf{Initialize: } $loaded \leftarrow 0, s \leftarrow 0, k \leftarrow 0$ $\forall$ new queries
    \newline    
    \Parallel
        \State \texttt{TRANSPORT} $(i, j, k) \; \forall \;i \in [\max(1, loaded - n), s]$
        \If {$loaded \leq n$}
            \State \texttt{LOAD}
        \EndIf
    \EndParallel
    \State $loaded \leftarrow loaded + 1$
    \newline
    \Parallel
        \State \texttt{ROUTE} $(i, j, k)  \; \forall \;i \in [\max(0, loaded - n - 1), s - 1]$
        \State \texttt{STORE} $(s, j, k)$
    \EndParallel
    \newline
    \Parallel
        \State \texttt{TRANSPORT} $(i, j, k) \; \forall \; i \in [\max(1, loaded - n), s]$
        \If {$loaded \leq n$}
            \State \texttt{LOAD}
        \EndIf
    \EndParallel
    \State $loaded \leftarrow loaded + 1$
    \newline
    \Parallel
        \State \texttt{ROUTE} $(i, j, k)   \; \forall \;i \in [\max(0, loaded - n - 1), s]$
    \EndParallel
    \State $s \leftarrow s + 1$
\end{algorithmic}
\end{algorithm}

Similarly, we define the inverse of the operations as \emph{\texttt{UNLOAD} (L')}, \emph{\texttt{UNTRANSPORT} (T')}, \emph{\texttt{UNROUTE} (R')}, and \emph{\texttt{UNSTORE} (S')} respectively. Note that the gates for the last three operations are identical to their reversed counterparts (e.g. \texttt{TRANSPORT} and \texttt{UNTRANSPORT} are both a \texttt{SWAP} gate), but are conceptually different in their role.

Using these instructions, we provide step-by-step algorithmic descriptions of the query procedure. We decompose it into into the following subroutines: Alg.~\ref{alg:query} for the overall Fat-Tree QRAM procedure, Alg.~\ref{alg:load} for address loading and Alg.~\ref{alg:unload} for address unloading. The latter two can also be applied to BB QRAM, and are referenced by Alg.~\ref{alg:query}.

\begin{figure*}[t]
         \centering
         \includegraphics[width=0.98\linewidth]{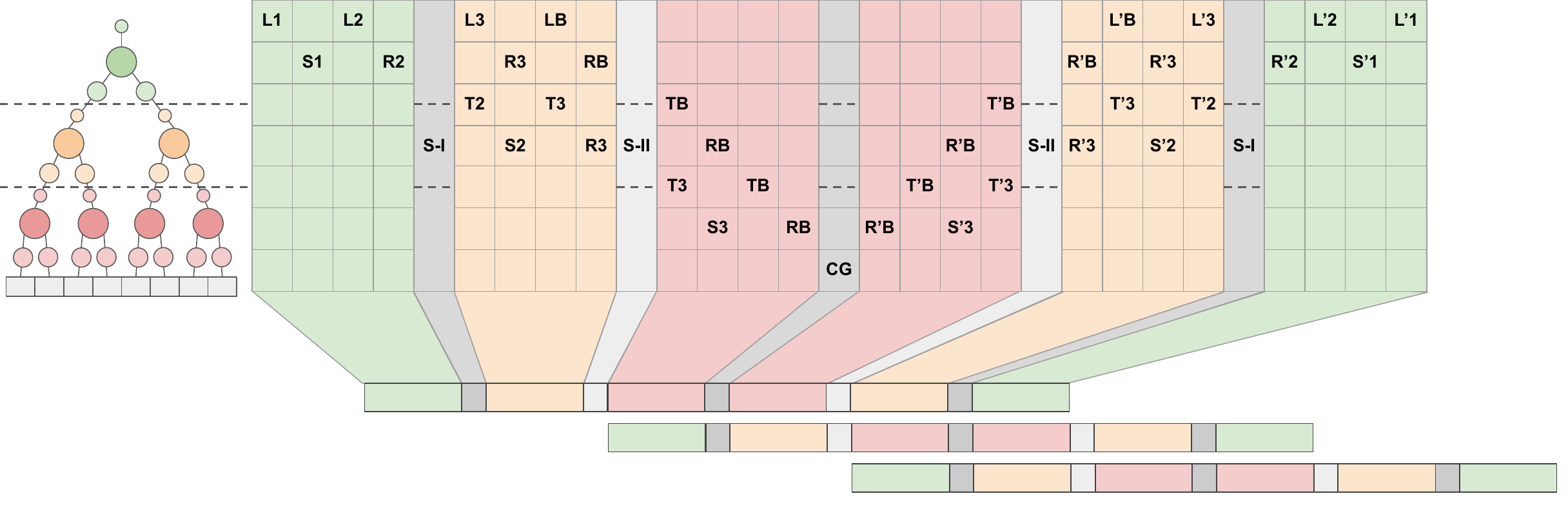}
         \caption{A step-by-step pipelining diagram for three capacity-8 queries using the instruction set defined in Sec. \ref{subsec:BBdetail}. Numbers in the operations refer to the information being moved by the operation with address qubits numbered 1 to 3 and $B$ denoting the bus (e.g. $S1$ represents storing the first address qubit). Columns indicate the circuit layer of the operation and rows denote the qubit in which the operation occurs. Similar to Fig. \ref{fig:fullpip}, colors indicate the conceptual QRAM $k$ being used and the type of SWAP-I/II. Note that the query pipelines all align and there is no conflicting usage of qubits.}
         \label{fig:detailbb}
\end{figure*}

\subsection{Proof of FIFO Scheduling Optimality }
\label{subsec:greedyproof}

\begin{algorithm}[t]
\caption{\textsc{Unload Layer}}\label{alg:unload}
\begin{algorithmic}[1]
    \State \textbf{Require: } $loaded$ as number of address qubits loaded
    \State \textbf{Require: }  $s$ as next address depth to be stored
    \State \textbf{Require: } $k$ as the current QRAM copy being used
    \State \textbf{Ensure: } runs only after data retrieval
    \newline
    \State $s \leftarrow s - 1$
    \Parallel
        \State \texttt{UNROUTE} $(i, j, k)   \; \forall \;i \in [\max(0, loaded - n - 1), s]$
    \EndParallel
    \newline
    \State $loaded \leftarrow loaded - 1$
    \Parallel
        \State \texttt{UNTRANSPORT} $(i, j, k) \; \forall \;i \in [\max(1, loaded - n), s]$
        \If {$loaded \leq n$}
            \State \texttt{UNLOAD}
        \EndIf
    \EndParallel
    \newline
    \Parallel
        \State \texttt{UNROUTE} $(i, j, k) \; \forall \;i \in [\max(0, loaded - n - 1), s - 1]$
        \State \texttt{UNSTORE} $(s, j, k)$
    \EndParallel
    \newline
    \State $loaded \leftarrow loaded - 1$
    \Parallel
        \State \texttt{UNTRANSPORT} $(i, j, k) \; \forall \;i \in [\max(1, loaded - n), s]$
        \If {$loaded \leq n$}
            \State \texttt{UNLOAD}
        \EndIf
    \EndParallel
\end{algorithmic}
\end{algorithm}

Using a greedy exchange proof, we show that for Fat-Tree QRAM, FIFO scheduling is optimal regarding overall query latency for both offline and online cases.

Consider a scheduling of queries $q_1, q_2, ..., q_n$ which are all requested at times $t_1, t_2, ..., t_n$ respectively. Let $s_i$ also denote the time that $q_i$ begins computing ($t_i \leq s_i$ since we can only start a query after it is requested and $s_i < s_j$ for all $i \leq j$) and $L_i = (s_i + T) - t_i$ denote the latency of $q_i$ where $T$ is the amount of time it takes to process a query (constant across all queries).

Suppose there is some optimal solution that does not follow our greedy FIFO scheduling. That is, there must exist two consecutive queries $q_x$ and $q_{x + 1}$ such that $t_{x + 1} \leq t_x$ (i.e. we schedule $q_x$ before $q_{x + 1}$ even though it is requested later). We show that swapping $q_x$ and $q_{x + 1}$ so they are scheduled in the order of their request times will result in a latency $L'$ no worse than the optimal (i.e. $L' \leq L$). We have that the total latency in the optimal scheduling is $L = \sum_{i} L_i = L_x + L_{x + 1} + \sum_{i \neq x, x + 1} L_i$.

If we swap $q_x$ and $q_{x + 1}$, the other queries' latencies will not change meaning $\sum_{i \neq x, x + 1} L'_i = \sum_{i \neq x, x + 1} L_i$. We can start $q_x$ at time $s_{x + 1}$ and $q_{x + 1}$ at time $s_x$ as the QRAM is available during both those times (otherwise it would not be available in the original scheduling) and the queries are still only started after being requested ($t_{x + 1} \leq t_x \leq s_{x} < s_{x + 1}$). This results in new latencies of $L'_x = (s_{x + 1} + T) - t_x$ and $L'_{x + 1} = (s_x + T) - t_{x + 1}$. It is easy to show that $L'_x + L'_{x + 1} = L_x + L_{x + 1}$ by rearranging the terms. Thus, $L = (L_x + L_{x + 1}) + \sum_{i \neq x, x + 1} L_i = (L'_x + L'_{x + 1}) + \sum_{i \neq x, x + 1} L'_i = L'$.

If we continually swap all pairs of such $q_x$ and $q_{x + 1}$ where $t_{x + 1} \leq t_x$, we can incrementally transform the optimal solution into our greedy FIFO scheduling. Since at each step our latency is no worse than before, the final latency of our FIFO scheduling is no worse than the optimal solution, making it another optimal solution as well. This completes the proof that a FIFO scheduling always minimizes the total latency.

\clearpage

%%
%% The next two lines define the bibliography style to be used, and
%% the bibliography file.
\bibliographystyle{ACM-Reference-Format}
\balance
\bibliography{refs}

%%
%% If your work has an appendix, this is the place to put it.

\end{document}